\documentclass[10pt,journal]{IEEEtran}

\usepackage{cite}
\usepackage{amsmath,amssymb,amsfonts}
\usepackage{algorithmic}
\usepackage{graphicx}
\usepackage{textcomp}
\def\BibTeX{{\rm B\kern-.05em{\sc i\kern-.025em b}\kern-.08em
    T\kern-.1667em\lower.7ex\hbox{E}\kern-.125emX}}

\usepackage{subfigure}
\usepackage[english]{babel}
\usepackage[utf8]{inputenc}
\usepackage[T5,T1]{fontenc}
\usepackage{enumerate}
\usepackage{soul}

\usepackage{indentfirst}
\usepackage{setspace}
\usepackage{units}
\usepackage{physics}
\usepackage[capitalize]{cleveref}

\begin{document}

\title{Quantum Error Correction via\\ Noise Guessing Decoding}

\author{Diogo Cruz, Francisco~A.~Monteiro,~\IEEEmembership{Member,~IEEE}, Bruno C. Coutinho
\thanks{This work was supported in part by the European Union’s Horizon 2020 Research and Innovation Program through the Project Quantum Internet Alliance (QIA) under Grant 820445; and in part by Fundação para a Ciência e Tecnologia / Ministério da Ciência, Tecnologia e Ensino Superior (FCT/MCTES), Portugal, through national funds and when applicable co-funded EU funds under projects UIDB/50008/2020 and 2022.05558.PTDC. Diogo Cruz was supported by FCT scholarship UI/BD/152301/2021.}

\thanks{D. Cruz is with Instituto de Telecomunicações, and IST, Universidade de Lisboa, Lisbon, Portugal, e-mail: \{diogo.cruz@lx.it.pt\}.}

\thanks{F. A. Monteiro is with Instituto de Telecomunicações, and ISCTE~-~Instituto Universitário de Lisboa, Lisbon, Portugal, e-mail: \{francisco.monteiro@lx.it.pt\}.}

\thanks{B. C. Coutinho is with Instituto de Telecomunicações, Lisbon, Portugal, e-mail: \{bruno.coutinho@lx.it.pt\}.}
}

\maketitle

\begin{abstract}
Quantum error correction codes (QECCs) play a central role in both quantum communications and quantum computation. Practical quantum error correction codes, such as stabilizer codes, are generally structured to suit a specific use, and present rigid code lengths and code rates. This paper shows that it is possible to both construct and decode QECCs that can attain the maximum performance of the finite blocklength regime, for any chosen code length when the code rate is sufficiently high.
A recently proposed strategy for decoding classical codes called GRAND (guessing random additive noise decoding) opened doors to efficiently decode classical random linear codes (RLCs) performing near the maximum rate of the finite blocklength regime. By using noise statistics, GRAND is a noise-centric efficient universal decoder for classical codes, provided that a simple code membership test exists. These conditions are particularly suitable for quantum systems, and therefore the paper extends these concepts to quantum random linear codes (QRLCs), which were known to be possible to construct but whose decoding was not yet feasible. By combining QRLCs and a newly proposed quantum-GRAND, this work shows that it is possible to decode QECCs that are easy to adapt to changing conditions. The paper starts by assessing the minimum number of gates in the coding circuit needed to reach the QRLCs' asymptotic performance, and subsequently proposes a quantum-GRAND algorithm that makes use of quantum noise statistics, not only to build an adaptive code membership test, but also to efficiently implement syndrome decoding.
	
\end{abstract}

\begin{IEEEkeywords}
	 GRAND, ML decoding, quantum error correction codes, short codes, syndrome decoding.
\end{IEEEkeywords}

\maketitle
\section{Introduction} 
Quantum technologies have the potential to revolutionize industries across the globe. By harnessing the power of quantum mechanics, technologies such as quantum computing, quantum simulation, quantum communication, and quantum sensing can be used to solve a wide range of hard problems in different areas, such as in the pharmaceutical, materials science, cryptography, machine learning, logistics optimization, finance, energy, and aerospace sectors \cite{alexQuantumTechnologiesReview2021,luckowQuantumComputingIndustry2021,martinQuantumTechnologiesTelecommunications2021a}. Nevertheless, for these technologies to be viable, significant progress needs to be made in quantum error correction, to reduce the error rates of these quantum systems to tolerable levels.

Implementing quantum computers and other devices bearing quantum memories will require quantum error correction of the stored quantum states \cite{devitt_quantum_2013,roffe_quantum_2019}. Moreover, a number of applications, such as quantum key distribution, distributed quantum computing, or the connection of different quantum sensors, require the existence of quantum networks and quantum repeaters in order to sustain end-to-end quantum connections between several devices \cite{dahlberg_link_2019, quantum_internet, coutinho_robustness_2022,Santos2023, bugalho2023, caleffi2018quantum}. Quantum states are notoriously sensitive to noise, so quantum error correction must be employed when establishing entanglement or for preserving quantum states in memory.
Quantum gates still exhibit very high error rates, even though they have fallen in the last decade from ${\approx}10^{-2}$ \cite{faulttolerance} to ${\approx}10^{-3}$ \cite{GoogleQ_Nature_2023}).
Thus, it is paramount to have scalable quantum error correction codes (QECCs), meaning that errors can be exponentially corrected while the encoding/decoding complexity increases sub-exponentially with the number of qubits in the codewords. A very recent breakthrough toward this objective was communicated in \cite{GoogleQ_Nature_2023}, showing experimental results for a surface code that seem to pave the way for practical and scalable QECCs.
The problem of quantum error correction has been looked at in\cite{Autoencoders_quantum_2023}
in a broader prospective than the traditional one of designing a fixed code. The authors proposed using quantum neural networks that work as autoencoders that optimally adapt the design of the QECC to the existing noise statistics. Note that making use of noise statistics is also a central idea of the proposal in the present work. QECCs designed for a particular type of channels, where the statistics of the errors show that one type of quantum errors is the most likely one, appeared in \cite{prevalent_error_types_JSAIT_2020}. 
\cite{Gullans_Krastanov_Huse_Jiang_Flammia_2021} presented an encoding method for quantum random codes and analyzed its performance.
\cite{tremblayFiniterateSparseQuantum2023} performed a similar analysis, but using a code construction akin to low-density parity-check codes (LDPCs) and based on constraint satisfaction. The works in \cite{Gullans_Krastanov_Huse_Jiang_Flammia_2021} and \cite{tremblayFiniterateSparseQuantum2023} are mainly concerned with the efficiency of the \textit{encoding} process, and both consider a decoder for an erasure channel, rather than a Pauli decoherence channel, which is considered in our work.
Furthermore, important findings for QECCs designed for the deletion channel (i.e., when a qubit is removed without notice from a stream of qubits) have been communicated in \cite{deletion_ISIT_2020}.

\subsection{Short linear codes: a historical context}
\label{sec:short_hist}

Many ideas for designing QEEC can be traced down to the developments made in classical error correction, and in \cite{Babar_Hanzo_2019} one can find an exhaustive list of those connections.
The concept of error correction appeared in tandem with the one of channel capacity in the pioneering work of Claude Shannon \cite{shannon}, who proved that the capacity of a channel could be attained by using a uniform-at-random code that maps any $k$ bits of information onto codewords $n$ bits long.
This mapping should be random, in the sense that each of the $2^k$ codewords should be assigned to one of the $2^n$ possible words, and should be made according to a uniform distribution.
Shannon's random code construction provably reaches capacity when the length of the codewords tends to infinity ($n \rightarrow \infty$) and the decoder picks, among all possible codewords, the most likely one, by applying maximum-likelihood (ML) decoding with the received observation and the codewords' prior distribution. Nevertheless, Shannon's codes are too complex and have never been used in practice because that would imply storing all the $2^k$ codewords at the transmitter side, and applying ML decoding at the receiver, which is an NP-hard problem. 

Classical random linear codes (RLCs) offer a solution to the codewords' storage problem, given that only the \textit{random} generation matrix needs to be stored. Moreover, they are known to achieve capacity in the binary symmetric channel \cite{shannon,Gallager_1973}. Although an exhaustive search among all the $2^k$ codewords of a RLC remains prohibitively complex, ML decoding is known to be attainable by other means. An exhaustive search entails searching through all the $2^k=2^{nR}$ codewords when the code rate is $R\leq 1/2$, however, when $R>1/2$, it is preferable to look at the $2^{(n-k)}=2^{n(1-R)}$ syndromes and perform syndrome-based decoding, which also delivers the ML decoding performance. In short, as it is well known, the complexity of an exhaustive search approach amounts to $2^{n \min(R, 1-R)}$ \cite{Coffey_Goodman_90}.
\\
It is known that the computational complexity of decoding \textit{any} (deterministic or random) linear block code (LBC) can be greatly reduced by using set information set decoding algorithms (see \cite{Coffey_Goodman_90} and references therein), or by having a trellis representation of the LBC and running the Viterbi algorithm on that trellis. The latter technique has been known since \cite{Wolf_1978}, when Wolf showed how to construct a trellis representation of a LBC using its parity-check matrix. However, as described in \cite{Kschischang_Sorokine_TIT_1995}, research on trellis decoding of LBCs was not very active until the paper by Forney (and in particular its appendix) came out \cite{Forney_TIT_1998}. In \cite{Kschischang_Sorokine_TIT_1995}, Kschischang and Sorokine provided a (arguably) more elegant manner of constructing the trellis of a LBC by using the spans of the so-called atomic codewords, which is a tool to find the \textit{unique} minimum trellis of a LBC. A measurement of the complexity of the resulting trellis is the maximum number of states existing across all its sections, and that number can be shown to be $s_{\max}=2^{\min(k, n-k)}$. This corresponds to the bound given by Wolf \cite{Wolf_1978} and also found in \cite{ShuLin_TIT_1993, Kschischang_Sorokine_TIT_1995}. Forney's work was, in fact, much broader, unlocking the trellis representation of lattice codes (of which LBCs are a particular case), and therefore permitting ML lattice decoding of specially designed lattices \cite{Amir_thesis_1997}, and also approximate ML decoding of random lattices \cite{Monteiro_Kschischang_11}.
\\
The path taken by coding theory (and coding practice) in the following years focused on finding practical codes that would achieve Shannon's capacity; regardless of the family of codes being studied, the efforts were concentrated on long codes (with large $n$) and typically not having a very high rate. This may explain why the work on trellis decoding of LBCs (summarized in \cite{Honary_book_97,ShuLin_book_98}), as well as the various information set decoding algorithms, have been overlooked as a possible low-complexity ML decoding solution for RLCs. All these decoding approaches deal with lists of size $2^{(n-k)}$: being it the number of syndromes, or the number of coset representatives described by the trellis of the code. While syndrome decoding would involve storing all those syndrome pairs and the associated coset representative in the so-called \textit{standard array}, a ML Viterbi trellis decoder implies storing a trellis with the same number of paths (each of which associated with a coset representative). When the redundancy of the code (i.e., $n-k$) is large, the memory requirements for syndrome decoding become too large, and in the case of trellis decoding, the trellis comprises too many states, in both cases limiting the real-world use of capacity-achieving RLCs.

Motivated by both power- and complexity-limited devices in wireless communications, and also due to the goal of reducing end-to-end coding/encoding latency, research only recently shifted to the finite-blocklength, with an emerging interest for short codes and high rates, eventually culminating in a practical method for decoding RLCs dubbed guessing random additive noise decoding (GRAND) algorithms \cite{duffy_capacity-achieving_2019, Duffy_2022, grand_mit, Patent_2020a, Patent_2020b, Patent_2021}.
At its core, GRAND is focused on guessing the noise that corrupted the transmitted codeword, rather than exhaustively going through all the possible codewords in order to find the one that fits the ML criterion, and is proven to still lead to ML decoding \cite{duffy_capacity-achieving_2019}. GRAND is a \textit{universal} decoder, enabling the decoding of \textit{any} block code of moderate length or high code rate, when dealing with low entropy noise. The only requirement is that a practical membership test exists to assess whether some word belongs to the codebook or not. The decoding complexity is measured by the average number of membership tests needed until a valid codeword is found. Given that the space of possible words has size $2^n$, and that there are $2^k$ valid codewords, that average number of attempts until one finds a codeword is given by the ratio $\frac{2^n}{2^k}=2^{n-k}=2^{n(1-R)}$. Therefore, that number can be low, provided that the codeword are short or that the code rate is high, which is particularly attractive for quantum setups, where the number of available qubits is modest.

Although ML decoding could be implemented for the particular case of low-redundancy RLCs via the above-listed techniques, one would still face memory constraints. Contrary to information set decoding, syndrome decoding, or trellis-based decoding, GRAND does the decoding on-the-fly, dispensing any type of storage besides the parity-check matrix of the RLC (or any particular LBC). Note that computing a minimal trellis of an RLC is a non-negligible preprocessing that is needed each time the RLC changes. Additionally, GRAND can outperform those techniques (designed for memoryless channels) in the case of non-independent and non-identically distributed noise having memory \cite{TComm22}. In fact, any prior information about the structure of the noise can lead to performance improvement or to further complexity reduction. For example, incorporating the geometry of the used symbol constellation can significantly reduce GRAND's complexity \cite{Chatzigeorgiou_Monteiro_2023}.

\subsection{Scope and Contributions}
Due to the technical difficulties in manipulating qubits, the error correction codes applied to qubit packets in quantum communication links or quantum memories are necessarily ones with short codewords, and for that reason, quantum error correction is a good fit for GRAND-inspired strategies. Since all quantum operations are required to be unitary, the construction of quantum random linear codes (QRLCs) is not as straightforward as in classical RLCs.
This difficulty in adapting classical codes to the quantum setting is typical; it also took an extensive effort to find quantum LDPC codes whose properties matched their classical counterparts, and these issues continue to be the focus of intensive research \cite{breuckmannQuantumLowDensityParityCheck2021a, LDPC_quasi-cyclic_stabilizer_TCOM_2018, Wang_TCOM_23}. Similar issues plagued the development of quantum turbo codes \cite{poulinQuantumSerialTurbo2009}. Furthermore, to borrow the performance guarantees existing in the classical setting, many of these adaptations amount to Calderbank-Shor-Steane (CSS) codes, which have suboptimal channel capacity \cite{calderbankGoodQuantumErrorcorrecting1996,steaneMultipleparticleInterferenceQuantum1996}.

Fortunately, for QRLCs, simple (randomized) constructions exist \cite{brown_short_2013, Gullans_Krastanov_Huse_Jiang_Flammia_2021}, leading to codes whose performance is primarily conditioned by the number of quantum gates used as building blocks for the coding circuit. For codewords with $n$ qubits, and in a setup with all-to-all connectivity (i.e., when any qubit can be directly entangled with any other), $\order{n \log^2 n}$ two-qubit random Clifford gates and $\order{\log^3(n)}$ circuit depth has been shown to lead to codes with a reasonable Hamming distance \cite{brown_short_2013}.
More recent results have expanded the capabilities of QRLCs for the case of limited qubit connectivity, and eased the gate requirements \cite{Gullans_Krastanov_Huse_Jiang_Flammia_2021}. In general, the more gates used, the more the code approaches a ``truly'' random QRLC. As part of the encoding for stabilizer codes, they may also constitute a vital piece in quantum error correction for quantum computing, quantum networks, and quantum memories. However, these codes have suffered from some of the same disadvantages as classical RLCs, as they have been too difficult to decode in practice.
In fact, to the authors' best knowledge, only recently has been presented in \cite{darmawanLowdepthRandomClifford2022} a practical decoding technique, based on tensor networks, applicable specifically to low-depth circuits with one-dimensional connectivity. They numerically verify that their code reaches the hashing bound \cite{wilde_2017}, but do not prove it mathematically. The recent work in \cite{darmawanLowdepthRandomClifford2022} does consider a Pauli decoherence channel, however the results are presented using different variables and parameters, difficulting a direct comparison with our proposal.
We note that, from a practical perspective, our model has some limitations. We consider all-to-all qubit connectivity, noiseless encoding circuits, as well as a noiseless syndrome extraction procedure. We also consider noiseless quantum gates. These assumptions are regularly used to analyze a code's performance \cite{landahlFaulttolerantQuantumComputing2011}, and they are used similarly in other results in the literature \cite{Gullans_Krastanov_Huse_Jiang_Flammia_2021,darmawanLowdepthRandomClifford2022,tremblayFiniterateSparseQuantum2023}. These limitations are further addressed in \cref{sect:conclusions}.

Interestingly, research on trellis decoding has recently also begun being studied as a possible practical way of implementing ML decoding for any quantum stabilizer code over a finite field of prime dimension \cite{Sabo_GaTech_2022}. As in the case of the trellis decoding of classical RLCs (mentioned in Section \ref{sec:short_hist}), a trellis-based approach may also open doors to the decoding of QRLCs.

The present paper extends the noise-decoding concept to quantum systems, allowing one to tackle the decoding of QRLCs for the first time in the typical Pauli decoherence channel. Meanwhile, results on the use of guesswork in the decoding of designed families of quantum codes have appeared in \cite{Hanzo_Access_2023}. The concept of universal guessing-based decoding was tested there with non-random codes, i.e., quantum codes that are known and hold a design structure that fits particular decoding methods.

Our work considers a system model with perfect encoding and perfect measurements, focusing on correcting the errors that arise in the communication channel, which is a common model used in the literature \cite{Swathi_Access_2022, Hanzo_Access_2023}.
By making use of the key features of GRAND, this work shows how to take advantage of QRLCs to construct stabilizer codes that are shown, via semi-analytical simulation, to be near-capacity achieving and decodable in practice, leading to quantum error correction codes able to cope with large channel error rates, even at reasonably high code rates.

As it is a universal decoding method, the proposed quantum-GRAND (QGRAND) constitutes a general approach to QEC that is amenable to any code length size $n$ and a sufficiently high code rate $R$. Furthermore, it is highly adaptable to on-the-fly changes depending on noise statistics, and is likely suitable to scenarios where less than all-to-all connectivity exists.
For these reasons, the combination of QRLCs and QGRAND stands out from the well-established QECCs. This combination is potentially of high importance to future implementations of quantum error correction for quantum computing and quantum networks, as recently highlighted in \cite[sec. VI.B]{Access_Hanzo_Sep2023}.

The literature on the use of GRAND to decode classical RLCs is quite recent and it often directly compares the decoding complexity of RLCs via GRAND with the one of ML decoding. The historical context (given in \cref{sec:short_hist}) brought to attention several other approaches that enabled the decoding of RLCs. That account is further complemented in Section \ref{sect:CRLC}. Besides that, the main contributions of the paper are i) the analysis of the requirements for the construction of a good QRLC, including the implications to the associated encoding complexity, measured by the number of gates in the encoding circuit, and ii) the proposal of a quantum-GRAND approach that enables the numerical assessment of the performance of QRLC's for the first time, and measuring how close those results are from the optimal performance, which is analytically derived.

\subsection{Organization and Notation}
The paper is organized as follows: Section II presents a brief introduction to classical RLCs, and introduces GRAND in the classical domain.
Section III provides some quantum error correction basics, and Section IV details the approach taken by stabilizer codes.
Section V presents some theoretical and numerical results (using a semi-analytical framework) for QRLCs, and the noise model considered. Section VI describes how to construct a stabilizer code from this quantum RLC encoding.
These ideas are later combined in Section VII, where QGRAND is showcased. The numerical results are displayed in Section VIII, and a final discussion appears in Section IX.

For the sake of clarity, we opted to denote matrices and vectors in non-bold, as that is the traditional notation in the field of quantum signal processing, with the exception of the classical vector $\vb s$ and the zero vector $\vb 0$.

It should be noted that \textit{ error rate} is used throughout the paper in the same sense that \textit{bit error rate} is used in classical communications, meaning that herein the term is used with respect to qubits \textit{after} their discretization. The literature on QECCs often calls it \textit{error probability} instead, and rather associates the error rate with the continuous errors of the \textit{physical} qubits \cite{Terhal_2015}.

\section{Toward Shannon's Codes: Classical RLCs} \label{sect:CRLC}
In 1948, Shannon proved \cite{shannon} that (classical) random error correcting codes with codewords of length $n$ and having $2^k$ valid codewords randomly chosen out of the $2^n$ possible words are able to achieve the capacity of the Gaussian noise channel, as $n\rightarrow \infty$. Even so, randomly selecting the codebook members leads to: i) a storage problem, given that all codewords would have to be stored both at the encoding side and at the decoding side, and ii) a decoding complexity problem, given that, when applying ML decoding, a corrupted codeword needs to be compared with all codewords in the codebook.

The storage problem posed by Shannon's construction (to generate uniform-at-random codes) has been overcome by RLCs, of rate $R=k/n$, because, as in the case of any LBC, its generator matrix constitutes a very short description of the code.
Without the size constraints of many families of classical structured codes, that impose constraints both on the admissible codeword lengths and code rates, RLCs can be constructed with any size and rate, and having those degrees of freedom is a major practical advantage for most engineering applications.

\subsection{Decoding Classical RLCs by Noise Guessing} \label{sect:ClassicalGRAND}

A paradigm change in the decoding of classical codes recently occurred for codes with short codewords or having low redundancy (i.e., high rate). GRAND \cite{duffy_capacity-achieving_2019,grand_mit, Patent_2020a, Patent_2020b, Patent_2021} fundamentally shifts the role of detection from searching \textit{codewords} in the codebook to searching for the \textit{error pattern} that took place.

Let us assume that a received block $Y=X \oplus E_i$ has been affected by some error pattern $E_i$, which could have occurred with probability $p_i$. The statistics $\mathcal{N}$ of the possible error patterns are known, and the $N$ candidate error patterns $E_i$ are ordered from the most likely ($i=1$) to least likely ($i=N$). GRAND consists of subtracting the error pattern $E_i$ from $Y$, starting from $i=1$ and progressing down the list in $\mathcal{N}$. If $X = Y \ominus E_i$ is a valid codeword, then $X$ is accepted as the decoded data. Otherwise, it moves to the next error pattern $E_{i+1}$ and the process is repeated. Here, it is important to note that having a tool to test if $X$ is a codeword (i.e., a \textit{membership test} \cite{duffy_capacity-achieving_2019}) is absolutely necessary.
If our noise statistics $\mathcal{N}$ encompasses all possible error patterns, it is guaranteed that a valid codeword $X$ is found at some iteration.
As proven in \cite{duffy_capacity-achieving_2019}, for noise with Shannon entropy $H(\mathcal N)$ (or entropy per bit $h(\mathcal N)\triangleq H(\mathcal N)/n$) and a code rate $R=k/n$ below capacity (i.e., $R < 1-h(\mathcal N)$), the expected number of error patterns needed to be tested until finding the true noise pattern is $2^{n \min\qty{h_{1/2} (\mathcal N), 1-R}}$, for large $n$, where $h_{1/2}$ is the Renyi entropy of order $1/2$ \cite{Arikan}, per bit. GRAND works best when the noise entropy is low or the code rate is high. In that case, the set of $2^n$ words is densely populated by codewords, and therefore the number of tests can easily be much lower than $2^{nR}$, the expected number of tested patterns for standard ML decoding. Given the focus on decoding the noise, exploiting the prior knowledge of the noise statistics plays a central role in boosting the overall decoding performance of GRAND \cite{TComm22}.

GRAND performs ML decoding and can therefore provide capacity-achieving performance when decoding structured capacity-achieving codes or RLCs. For practical reasons, one may be interested in the most likely errors only, and the search may be abandoned after iterating through slightly more than the first $2^{nh (\mathcal N)}$ error patterns, and quit afterward. This variant is called GRAND with abandonment (GRANDAB) and still provides ML decoding.
In this manner, GRANDAB is also capacity-achieving, using many fewer iterations.

Although these results have been rigorously proven for uniform-at-random codes as $n\rightarrow \infty$, experimental applications using structured linear codes of small blocklength, and also RLCs in particular, have produced striking results \cite{Duffy_2021,An_Medard_Duffy_2021,solomon_soft_2020,An_Soft_2022, TComm22, Duffy_2022}, consistent with these theoretical guarantees.

It should be credited that very related ideas appear in \cite[pp. 220-227]{MacKay}, namely i) codeword guessing decoding based on the posterior probability of the noise, and ii) proving Shannon's capacity theorem for RLCs via counting arguments based on the number of distinct syndromes.

\section{Quantum error correction} \label{sect:QEC}

Quantum error correction attempts to restore a decohered quantum state, which has undergone some error, back to a quantum state encoding the original data \cite{devitt_quantum_2013, roffe_quantum_2019}. Qubits can undergo more types of errors than their classical counterparts, making QECCs in general more complex. Moreover, while classical use cases, such as wired communications or data storage, commonly offer error rates below $10^{-5}$, current quantum technologies need to be able to handle much higher error rates, similar to the ones encountered in uncoded classical wireless links (between $10^{-3}$ and $10^{-2}$) \cite{faulttolerance, GoogleQ_Nature_2023}. In addition to that, the construction and decoding of QECCs faces extra difficulties  \cite{djordjevic_quantum_2021}:
\begin{enumerate}[i)]
	\item the non-cloning theorem shows it is impossible to copy an arbitrary quantum state, thereby forbidding the use of copies as redundancy;
	\item quantum errors are continuous in nature, as the quantum state can be in any superposition of the basis states;
	\item any measurement performed on a state in superposition destroys that superposition, and thus its quantum information is lost.
\end{enumerate}

Surprisingly, it is possible to circumvent these issues and construct QECCs that reliably protect quantum data against errors. Just as in the classical case, QECCs work by encoding a quantum state composed of $k$ qubits into one with $n$ qubits, in such a way that errors can be detected and corrected.

Any quantum channel error $E$, discrete or continuous, acting on a quantum state, can be written as a linear combination of these Pauli matrices. For one qubit only, one has
\begin{equation}
	E = \alpha_0 I + \alpha_1 X + \alpha_2 Z + \alpha_3 Y,
\end{equation}
for some set of complex values $\alpha_i$. $I$ is a $2\times 2$ identity matrix and
\begin{equation}
	X=\mqty(0&1\\1&0),\quad Z=\mqty(1&0\\0&-1), \quad Y= iXZ,
\end{equation}
are Pauli matrices.
Since $Y = iXZ$, as long as QECCs can correct $X$ (bit-flip) and $Z$ (phase-flip) errors simultaneously and for each qubit, then they can also correct any arbitrary $n$-qubit error.

From this digitization, it can be shown \cite{djordjevic_quantum_2021} that a quantum error channel for $n$ qubits may be decomposed as a combination of Pauli strings
\begin{align}
	&P_n = e^{i\frac{\pi}{2}\theta}\; O_1 \otimes \ldots \otimes O_n,\notag\\
	\text{with }&\theta \in \qty{0,1,2,3},\quad O_i \in \qty{I, X, Y, Z}.
\end{align}
and $\otimes$ denotes the tensorial (or Kronecker) product. This description is named Pauli channel. 

The Clifford group is the set of unitary transformations that transform any Pauli string into a Pauli string.
Clifford circuits are built solely from unitaries of the Clifford group. Any such Clifford unitary can be implemented via composition of the simple 1- and 2-qubit unitaries
\begin{equation}
	\hspace{-1.5ex}H = \frac{1}{\sqrt{2}}\mqty(1 & 1 \\ 1 & -1), \sqrt{Z} = \mqty(1 & 0 \\ 0 & i),
	\text{CNOT} = \mqty( I &  0 \\  0 &  X).\label{eq:Clifford_gates}
\end{equation}
With these Clifford circuits, it is possible to construct an encoding for a QECC, in order to protect the desired quantum data. The backbone and inspiration for most quantum error correction methods are the so-called stabilizer codes \cite{GottesmanPhD}, and for that reason, this work will chiefly focus on these codes as the core machinery needed to produce random QECCs.

\section{Stabilizer codes} \label{sect:stabilizer_code}

\begin{figure}[t]
	\centering
	\includegraphics[width=1 \linewidth]{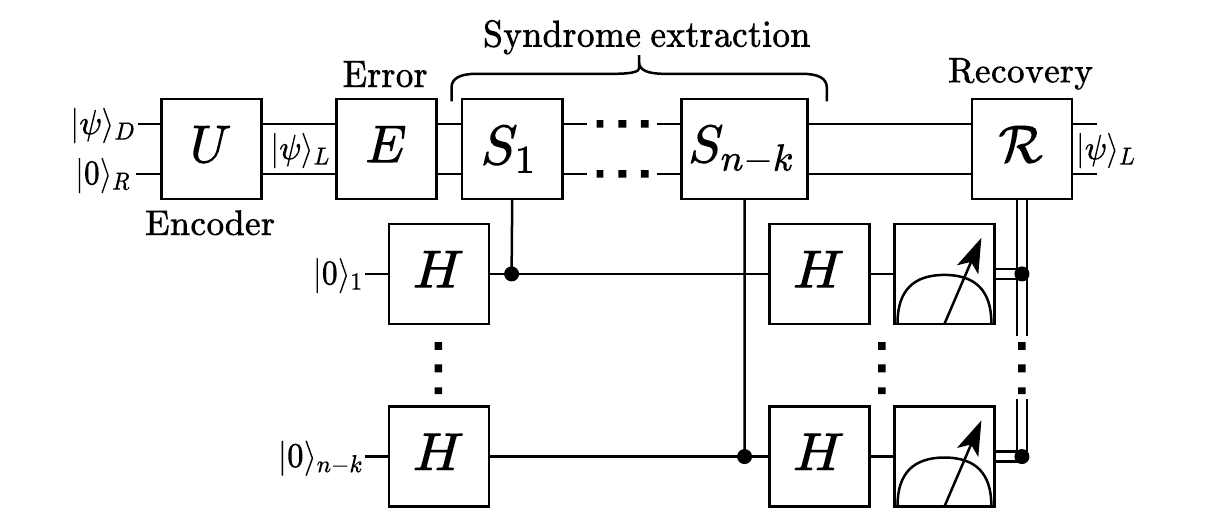}
	\caption{Quantum circuit for a $(n,k)$ stabilizer code, depicting the encoding and decoding process.}
	\label{fig:stab_code}
\end{figure}

A stabilizer $S_i$ of the quantum state $\ket \psi$ is any Pauli string that acts as the identity on $\ket \psi$, that is, $S_i\ket \psi = \ket \psi$. The stabilizer group $\mathcal S$ is the set of all such stabilizers. As there is a one-to-one correspondence between $\ket \psi$ and its stabilizer group, stabilizer circuits can be efficiently simulated using only Clifford unitaries (see \cite{gottesman_heisenberg_1998, aaronson_improved_2004}). The stabilizer group can be generated by some (non-unique) subset $\mathcal S_{\mathrm{min}}$, whose elements are the minimal stabilizers. These minimal stabilizers can be used to create a type of QECC, called a \textit{stabilizer code} (see Fig. \ref{fig:stab_code}). For simplicity, henceforth we will refer to the minimal stabilizers of stabilizer codes simply as stabilizers.

Starting with the $k$-qubit quantum data $\ket{\psi}_D$ and $(n-k)$ redundancy qubits in the state $\ket{0\ldots 0}_R$, which have been entangled by some Clifford unitary encoder $U$, then the resulting $n$-qubit logical state $\ket \psi_L$ (amounting to a codeword) must have $s\triangleq(n-k)$ stabilizers $\mathcal S_{\mathrm{min}}$, which must commute with each other. It also has $2k$ logical operators $\bar X_i, \bar Z_i$, for $1\leq i \leq k$, which act on $\ket \psi_L$ similarly to $X_i, Z_i$ on $\ket \psi_D$, and which can be combined to form a total of $L\triangleq2^{2k}$ logical operators.

Suppose that a Pauli error string $E$ affects $\ket{\psi}_L$, resulting in $\tilde{\ket{\psi}}_L = E\ket \psi_L$. If we conditionally apply a stabilizer to $\tilde{\ket{\psi}}_L$, controlled by an ancilla qubit in the state $\ket{+}= H\ket 0 = (\ket{0}+\ket{1})/\sqrt{2}$, and subsequently apply the Hadamard gate $H$ a second time (see \eqref{eq:Clifford_gates} \eqref{eq:Clifford_gates}), we obtain
\begin{align}
	&E\ket{\psi}_L \frac{\ket{0} + \ket 1}{\sqrt{2}}\\
	\stackrel{S_i}{\longrightarrow}\; & \frac{1}{\sqrt{2}}\qty[E\ket \psi_L \ket 0 + S_i E \ket \psi_L \ket 1]\\
	\stackrel{H}{\longrightarrow}\; & \frac{1}{2} \qty[(I+S_i)E\ket\psi_L \ket 0 + (I-S_i)E\ket \psi_L \ket 1].\label{eq:stabilizer_state}
\end{align}

Since both the error patterns and the stabilizers are Pauli strings, they must necessarily commute or anticommute. If they commute, the state in \eqref{eq:stabilizer_state} will be $E \ket \psi_L\ket 0$, otherwise, we will have $E \ket \psi_L \ket 1$. Consequently, the conditional application of each stabilizer provides us with 1 bit of information about the nature of the error. With $s$ stabilizers, we have a $s$-bit syndrome, allowing one to discern between $S \triangleq 2^{s}$ error patterns, while in the case of no error, one obtains the zero syndrome.
This information enables us to correct the quantum state by undoing the effect of the error. This recovery process $\mathcal R$ will consist of applying the inverse of the error pattern $E$ ascertained from the syndrome. Since all Pauli strings are both Hermitian and unitary, for a Pauli channel, we have $E^{-1} = E$.

\section{Quantum RLCs encoding}

The construction of classical RLCs mentioned in \cref{sect:CRLC} involves random generator matrices, whose only constraint is that they ought to be full-rank. Because quantum operators should be unitary, the construction of QRLCs is a rather more constrained problem. The generation of QRLCs is nevertheless possible: starting with a quantum state of $k$ qubits, to be encoded into $n>k$ qubits, \cite{brown_short_2013} presents a method of generating a random qubit encoding. One starts by randomly selecting Clifford unitaries from the $\mathcal C_2$ group (i.e., Clifford unitaries for 2 qubits). There are $|\mathcal C_2|=11\,520$ such unitaries, and all of them can be built by simple combinations of the Hadamard, CNOT, and phase gates, which have efficient physical implementations in virtually any quantum setting. After selecting these random unitaries from $\mathcal C_2$, one successively applies each of them to a random pair of qubits, taken from the set of $n$ qubits, assuming that all-to-all connectivity (between any of the $n$ qubits) is possible in practice. This assumption is required for the scaling results in \cite{brown_short_2013}, and is used in this work. However, it may be dropped for practical reasons, as the more recent results in \cite{Gullans_Krastanov_Huse_Jiang_Flammia_2021} suggest, and it is the subject of future work.
This process leads to an encoding unitary for our stabilizer code which, when applied to the initial $k$ qubits and $(n-k)$ extra $\ket 0$ qubits added, returns a $n$-qubit encoded quantum state. As shown in \cite{brown_short_2013}, as long as $\order{n \log^2 n}$ gates are used, with a circuit depth of $\order{\log^3 n}$, the construction leads to a highly performant $(n,k)$ code, and from \cite{Gullans_Krastanov_Huse_Jiang_Flammia_2021} it is already known that these complexity orders can be further lowered.

\subsection{Robust encoding}\label{sect:robust_encoding}

In this context, the QRLCs' performance will also depend on the noise statistics.
Nonetheless, in order to effectively protect our quantum data from noise in an efficient and practical manner, the noise statistics about the particular environment in question should be known. In the quantum setting, assuming we wish to correct the $N$ error patterns in the error set
\begin{equation}
	\mathcal E \triangleq \qty{E_i}_{i=0}^N, \text{ with }E_0\triangleq I,
\end{equation}
and we know the probability of them affecting our data to be $\mathcal P \triangleq  \qty{p_i}_{i=0}^N$, with $p_0\triangleq P(\text{no error})$, then our encoding and decoding procedure should have into account the noise statistics, denoted by the set of pairs
\begin{equation}
	\mathcal N \triangleq  \qty{(p_i, E_i)}_{i=0}^{N},
\end{equation}
with the error patterns ordered by decreasing probability, so that $p_i \geq p_{i+1}$, for all $i$.
For an encoding with $n$ qubits, using GRAND requires $N + 1 = 4^n$, with $\sum_i p_i = 1$, as $4^n$ is the total number of Pauli strings of $n$ qubits. If using GRANDAB, the unlikely errors may be disregarded, so a much lower $N$, and $\sum_i p_i \leq 1$, is assumed.

As the encoding is randomly chosen, to ascertain the code's performance, we may compute the syndromes associated with every error pattern in $\mathcal E$, and check whether the code is robust to all of them, in $\order{N}$ time. The code is robust if each $E_i\in\mathcal E$ has a unique syndrome or, for \textit{degenerate codes}, if $E_i$ shares a syndrome with $E_j$ and $E_iE_j\in \mathcal S$, that is, the combined action of the two errors does not change the quantum state.

Nonetheless, we may not require that all errors in $\mathcal E$ be correctable. For low entropy noise, some errors may be very unlikely to surface. If a random code happens to not correct these unlikely errors, it may still function adequately in the high probability case that more likely errors occur.

Classically, for a channel with input $X$ and output $Y$, its capacity is given by the maximum of the mutual information $I(X,Y)$ over the possible probability distributions of $X$. In the quantum case, for the case of stabilizer codes, the quantum capacity is given by the coherent information of the quantum channel \cite[sec. 7.6]{djordjevic_quantum_2021}, \cite[sec. 24.6.3]{wilde_2017}. For a Pauli channel, this capacity is associated with the so-called hashing bound \cite[sec. 24.6.3]{wilde_2017}, \cite{kingCapacityQuantumDepolarizing2003}. Using the results in \cite[sec. 24.6.3]{wilde_2017} we observe that, as long as
\begin{equation}
	n > k + H(\mathcal N),\label{eq:n_requirement}
\end{equation}
the code is able to correct the error patterns in $\mathcal N$ with high probability. This is a direct consequence of information-theoretic arguments based on set counting, and interpreting the entropy of the noise, $H(\mathcal N)$, as the number of bits needed to describe the noise, which establishes how many different noise patterns can occur. Note that, in theory, due to the possibility of degenerate error patterns, it is sometimes possible to surpass the hashing bound \cite{sutterApproximateDegradableQuantum2017,leditzkyUsefulStatesEntanglement2018}. Nonetheless, in \cref{sect:uniform_at_random} we motivate why degenerate errors have negligible effects in our setting, and consequently, in our numerical analysis in \cref{sect:results}, these error patterns are approximately considered to be uncorrectable.
Therefore, to obtain a near-capacity achieving code, we must only ensure that \eqref{eq:n_requirement} is satisfied, and that sufficient $2-$qubit gates are used to create the random encoding, as indicated previously, and as will be analyzed in \cref{sect:encoding}. This result is derived for large $n$, but similar results are applicable to lower $n$, as we see in the next section.

\subsection{Comparison with ideal random codes}\label{sect:uniform_at_random}
In order to assess the results obtained numerically, we may compare the QRLCs' performance with the expected performance of a simpler approximation. Similar techniques \cite{Gullans_Krastanov_Huse_Jiang_Flammia_2021} have been previously employed in the literature, since they allow us to better gauge the behavior of QRLCs, as it is too difficult to analytically estimate their performance.

As an ideal approximation, we consider a code that maps each of $N$ error patterns (plus the no error case) randomly, using an i.i.d. uniform distribution, to one of the $S$ syndromes. This approximation constitutes a more manageable mapping for analytical study, when compared with the mapping produced by a QRLC, since it does not present the linearity constraint, nor the added structure present in quantum codes. Nonetheless, since QRLCs may be thought of as a subset of the codes that this ideal approximation may generate, its performance may be considered to provide an upper bound for the QRLCs of interest.

For $N$ error patterns, plus the no-error case, let $u$ be the number of unique error syndromes associated with them.
In general, the average number of unique syndromes is
\begin{align}
	\ev{u}_{S,N+1} &= S\qty[1-\qty(1-\frac{1}{S})^{N+1}],\label{eq:ev(U)}
\end{align}
and the correctable error fraction is given by
\begin{align}
	f &= \frac{u_{S,N+1}}{N+1}.\label{eq:f_general}
\end{align}
If we wish to correct all $N$ error patterns (i.e., $f=1$), we must have $(N+1)$ distinct error syndromes, which happens with probability
\begin{align}
	P(f=1) &= \prod_{j=0}^{N}\qty(\frac{S-j}{S})\label{eq:f=1}\\
	&\simeq \prod_{j=0}^N e^{-j/S} = \exp(-\frac{N(N+1)}{2S}),
\end{align}
with the approximation achieved by the Taylor expansion of $e^x$.
Note that the probability of correcting absolutely all $N$ error patterns $P(f=1)$ may be very low, while still having $P(f>1-\epsilon)\simeq 1$, for $\epsilon\ll 1$. In fact, reminding that $S=2^{n-k}$, we have
\begin{align}
	P(f=1) \geq 1-\epsilon &\iff n \gtrsim k + \log_2\qty(\frac{N(N+1)}{2\epsilon})\label{eq:P_ineq}\\
	f \geq 1-\epsilon &\iff n \gtrsim k + \log_2\qty(\frac{N}{2\epsilon}).\label{eq:f_ineq}
\end{align}
These equations directly showcase the relation between the choice of $n$ for the encoding and the code's performance. Unlike \eqref{eq:n_requirement}, their derivation does not assume large $n$.

For $(n,k)$ QRLCs, each of the possible $4^{n}$ Pauli strings will be associated with one of the $S=2^{n-k}$ syndromes. Due to the linearity of the code, the strings will be distributed equally through the syndromes, so that each syndrome will have $2^{n+k} = SL$ associated Pauli strings. In fact, for $\vb s = \vb 0$, the associated strings will amount to all the combinations of the $S$ total stabilizers and the $L=2^{2k}$ total logical operators.

When creating a QRLC, it is possible to obtain a degenerate code. For an error pattern $E_i \in \mathcal E$, we obtain a degenerate code when $\exists i,j$ such that $i\neq j$ and $E_iE_j \in \mathcal S$. In these situations, $E_i$ and $E_j$ have the same effect on the encoded state and, consequently, can be treated as the same error. For two random error patterns $E_1, E_2$ with the same syndrome, the probability of $E_1E_2$ being a stabilizer is $\sim 1/L$.

For simplicity, we consider a code that can fully correct the error patterns in $\mathcal E$ a (good) degenerate code, while making no distinction between faulty codes that do or do not present some level of degeneracy.

\begin{figure}[t]
	\centering
	\includegraphics[width=0.95 \linewidth]{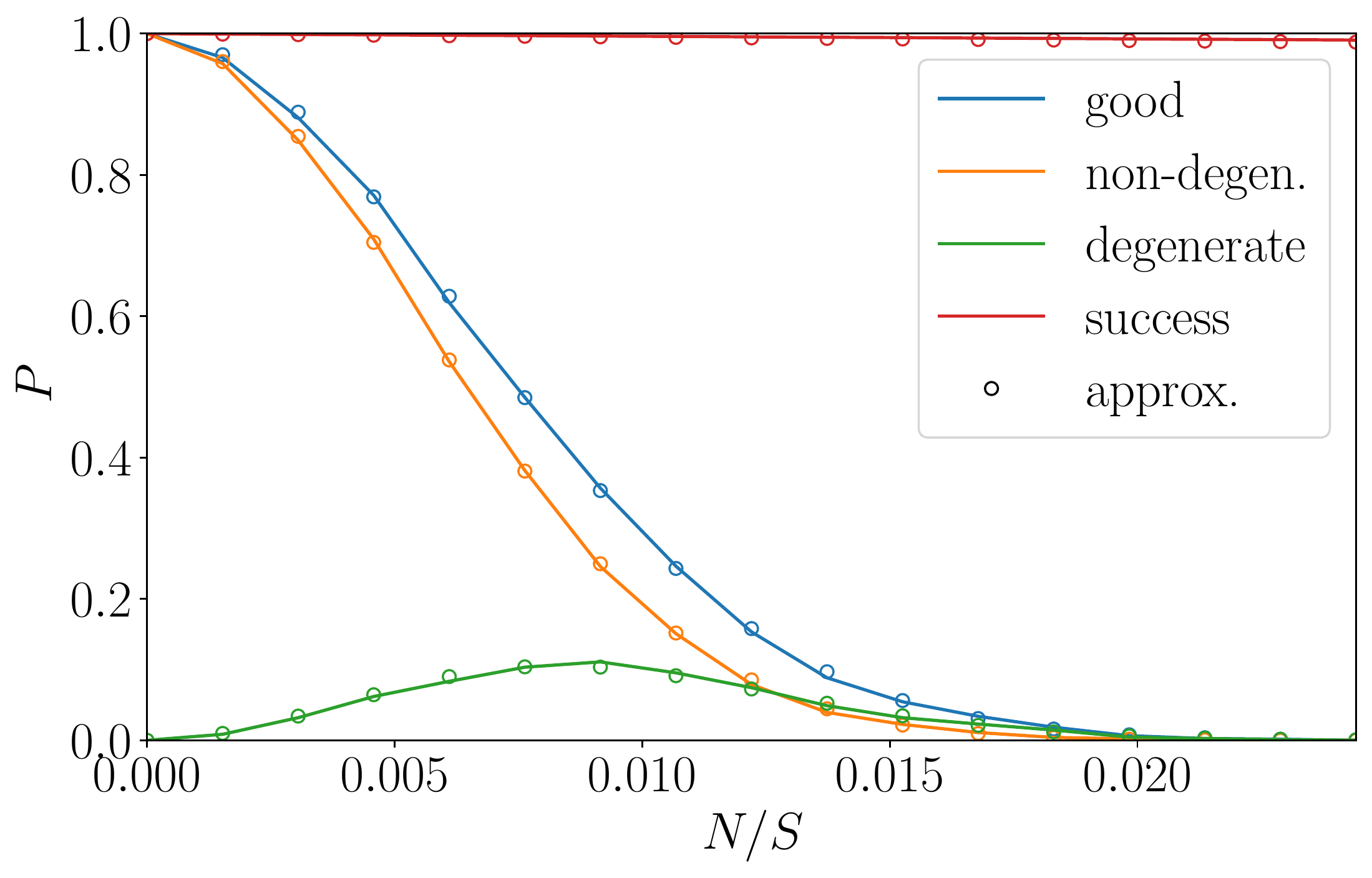}
	\caption{Probability of obtaining a ``good'' code (blue), that corrects all $N$ error patterns, vs. the ratio $N/S$, for a $(16,1)$ code, for the maximum noise entropy case. Even if obtaining a good code is unlikely, it still corrects almost all errors, so the probability of a successful correction is high (red). The dots indicate the theoretical approximations given by \eqref{eq:P_good}, \eqref{eq:f=1}, \eqref{eq:P_degenerate}, and \eqref{eq:f_general}, respectively, which closely match the observed behavior.}
	\label{fig:degenerate_codes}
\end{figure}

The probability of obtaining a ``good'' code, i.e., a code that is capable of correcting all the $N$ error patterns, can be estimated by considering the probability that either the syndrome of $E_iE_j$ differs from $\vb 0$ (non-degenerate case) or equals a stabilizer (degenerate case). As there are $\smqty(N+1\\ 2)$ such $E_iE_j$, this probability is given approximately by
\begin{align}
	P(\text{good}) &\simeq (P(\text{non-degen.}) + P(\text{degen.}))^{\smqty(N+1\\ 2)}\\
	& \simeq \qty[\frac{S-1}{S} + \frac{S}{4^n}]^{\smqty(N+1\\ 2)}\\
	& = \qty[1 - \frac{1}{S}\qty(1-\frac{1}{L})]^{\smqty(N+1\\ 2)}\\
	& \simeq \exp(-\frac{N(N+1)}{2S}\qty(1-\frac{1}{L})).\label{eq:P_good}
\end{align}
As the probability of obtaining a working non-degenerate code is approximately given by \eqref{eq:f=1}, the probability of obtaining a working degenerate code can be estimated by
\begin{equation}
	P(\text{degenerate}) = P(\text{good}) - P(\text{non-degenerate}).\label{eq:P_degenerate}
\end{equation}
Fig. \ref{fig:degenerate_codes} presents 2000 samples of $(16, 1)$ random codes. Most of the ``good'' codes are non-degenerate, so every error pattern has its distinct syndrome, but a small percentage may be degenerate, since $k=1$ is very low. As expected, the probability of obtaining a code that corrects all $N$ error patterns present quickly decreases with increasing $N$. Nonetheless, the probability that faulty codes will succeed in correcting any particular error is still high for larger $N$ (red). For the highest shown $N/S$, with $N/S=0.025$, $P(\text{success})=0.99$, while $P(\text{good})\simeq 0$. As noted previously, this large discrepancy between $P(\text{success})$ and $P(\text{good})$ stems from the fact that we may have $P(f>1-\epsilon) \simeq 1$ while also having $P(f=1)\simeq 0$.

\subsubsection{Bernoulli process} \label{sect:Bernoulli_process}

As an example, we may consider the noise model in which quantum errors consist of a Bernoulli process, with each individual qubit having a probability $p$ of suffering an error. The error can be of type $X, Y,$ or $Z$ (i.e., depolarizing noise), and the error types are identical and independently distributed (i.i.d.) for each qubit.

The number of distinct error patterns of weight $t'\leq t$ is
\begin{equation}
	B_t \triangleq  \sum_{t'=0}^t A_{t'},\qquad \text{with }A_t \triangleq  3^{t} \mqty(n\\ t).\label{eq:B_t}
\end{equation}
For $t'<t$, as error patterns of weight $t'$ are always more likely than error patterns of weight $t$, for this noise model, then whenever an error pattern of weight $t$ has the same syndrome as a weight-$t'$ error pattern, it cannot be corrected. Here, the degenerate scenarios are disregarded, since they are negligible for large $k$. The expected number of syndromes $M_t$ with a weight-$t$ error pattern as the most likely one mapped to them is $M_t = \ev{u}_{S, B_t} - \ev{u}_{S, B_{t-1}}$. The correctable error fraction $f$ of weight-$t$ error patterns that can be corrected will be, on average,
\begin{equation}
	f = \frac{M_t}{A_t}
	\simeq  \frac{S}{A_t}e^{-\frac{B_{t-1}}{S}}\qty(1-e^{-\frac{A_t}{S}}), \text{ for }S\gg 1.\label{eq:f}
\end{equation}
This expression enables a useful comparison for the performance of QRLCs, as shown in Fig. \ref{fig:relative_distance} (and later in figures \ref{fig:128} and \ref{fig:32}).

\subsection{Minimum number of random gates} \label{sect:encoding}

It is important to assess how deep (i.e., number of concatenated gates) the generator circuits need to be such that the QRLC they generate approximates the ideal performance.
In order to do so, we have implemented the methodology in \cite{brown_short_2013} and, for a varying number of gates, we measured the relative deviation 
\begin{align}
	\delta_f &\triangleq \qty(f_{\text{theory}}-f_{\text{exp}})/f_{\text{theory}},\\
	\delta_P &\triangleq P(f=1)_{\text{theory}}-P(f=1)_{\text{exp}},
\end{align}
where the $f$ stands for the fraction of weight-$t$ error patterns a code can correct. $f_{\text{theory}}$ amounts to the average correctable fraction coming from the approximation in \eqref{eq:f_general}, and $f_{\text{exp}}$ is the result of averaging the observed correctable fraction of 31 QRLC samples, for a given number of gates. Similarly, $P(f=1)_{\rm theory}$ is given by \eqref{eq:f=1}, and $P(f=1)_{\rm exp}$ is the ratio of the 31 samples that had $f=1$ for weight $t=1$. For the noise statistics, we assume a Bernoulli process, as previously described in \cref{sect:Bernoulli_process}.

\begin{figure}[t]
	\centering
	\subfigure[]{\includegraphics[width=0.95\linewidth]{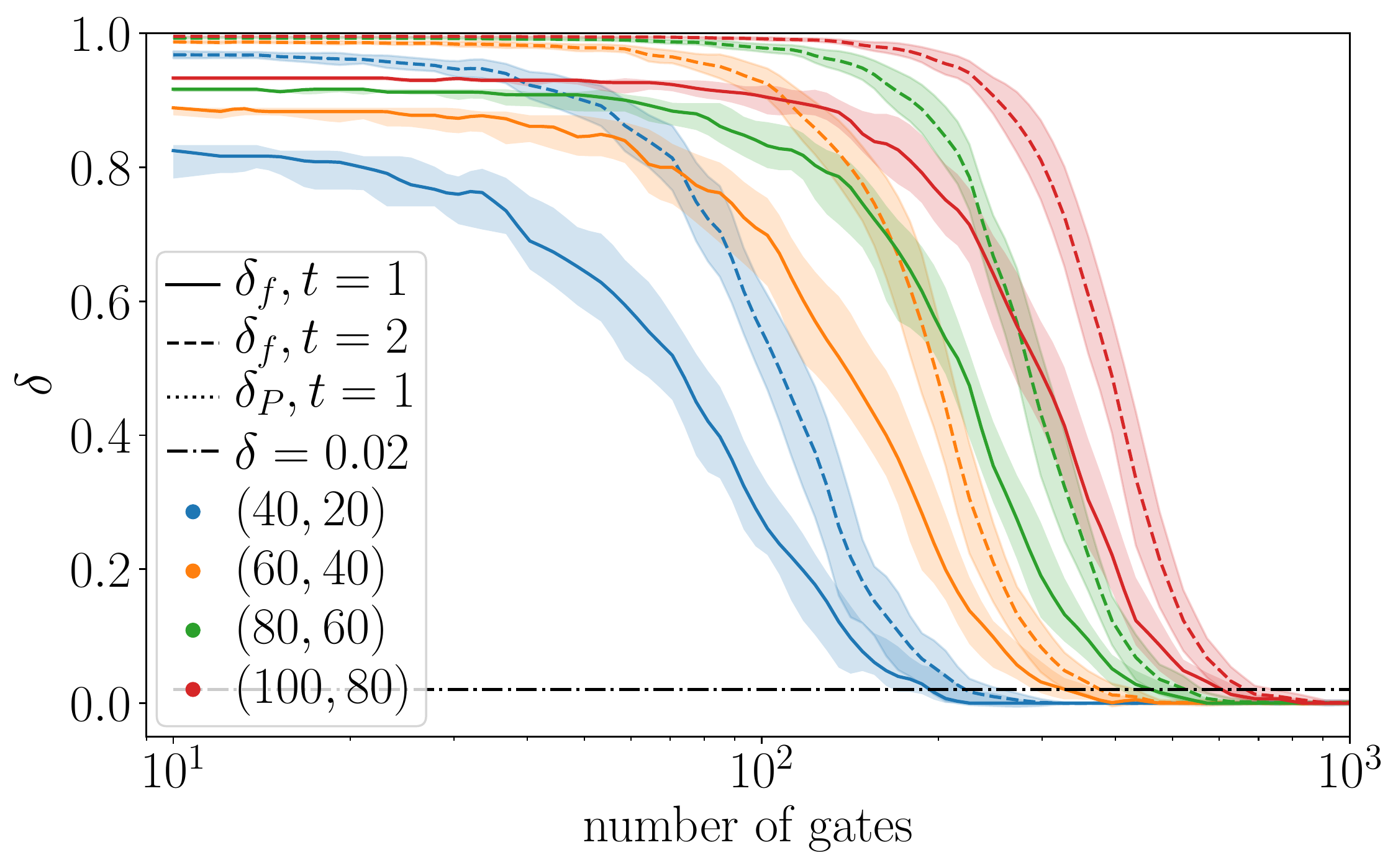}}
	\subfigure[]{\includegraphics[width=0.95\linewidth]{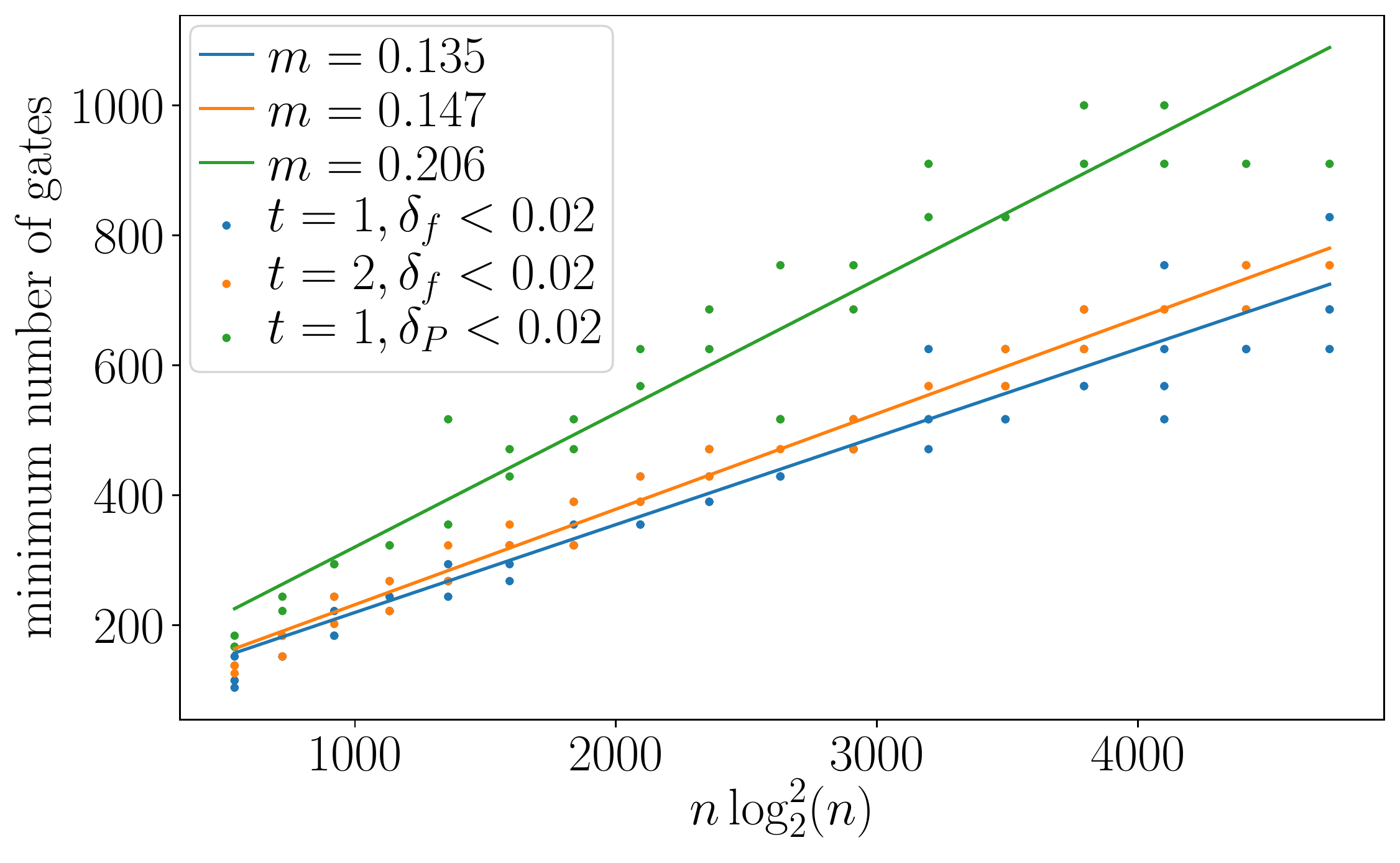}}
	\caption{(a) Relative deviation $\delta$, for different $(n,k)$ codes and error patterns of weight $t$. The shaded region indicates the $80\%$ confidence interval, obtained by constructing 31 random codes for each case presented. (b) Minimum number of gates to achieve a relative deviation $\delta < 0.02$, for varying $n\log_2^2 (n)$.}
	\label{fig:relative_distance}
\end{figure}

The simulation results presented in
Fig. \ref{fig:relative_distance} indicate that, as the number of gates used for the encoding increases, so do the codes' correction capabilities, eventually reaching ($\delta < 0.02$) those of the ideal random code. The codes' error correction capabilities start off worse for error patterns of larger weight, but converge to a similar minimum number of gates needed. We considered codes for several $n$ and $k$ to estimate this behavior. These results, shown in Fig. \ref{fig:relative_distance}, indicate that this minimum number of gates is given by $N_{\text{min. gates}}=m n \log_2^2(n)$. From \cite{brown_short_2013}, we expect the minimum number of gates to scale with $\order{n\log^2(n)}$. Although this theoretical result is only proved there for some codes, here we experimentally observe it for the mean $(n,k)$ code (with $m\simeq 0.13-0.15$), which is of more practical relevance, and for the case where we require that all $t=1$ error patterns must be correctable ($m\simeq 0.21$). Consequently, even for larger quantum codes with $n=128$, 1000 $\mathcal C_2$ Clifford gates should suffice to ensure that the resulting code has a very high performance for that rate and code length. Given that most of these gates will be applied in parallel, with depth $\order{\log^3 n}$, this approach enables the creation of high-capacity codes with low-depth quantum circuits.

\subsection{Determining the stabilizers} \label{sect:stabilizers}

The traditional approach in the construction of stabilizer codes is to start by defining the stabilizers, and then find the associated encoding process. The proposal in this paper uses a random code construction, and therefore the first step is the definition of the encoding circuit and only after one determines the stabilizers associated with that particular code.

The construction of a QRLC only involves Clifford unitaries, and therefore the encoding process $U$ is a stabilizer circuit. The Gottesman-Knill theorem indicates that stabilizer circuits with $n$ qubits can be simulated using only $\order{\text{poly}(n)}$ classical resources (see \cite{gottesman_heisenberg_1998, aaronson_improved_2004}). From this efficient classical simulation, the desired stabilizers may be extracted \cite{Gullans_Krastanov_Huse_Jiang_Flammia_2021}.
In particular, the $(n-k)$ ancilla qubits used for the encoding, in state $\ket 0$, start by having the associated minimal stabilizers $Z_i$, for $k+1\leq i\leq n$, since $Z_i \ket 0_i = \ket 0_i$. The classical simulation takes these starting stabilizers and modifies them through the encoding $U$, resulting in the desired minimal stabilizers $S_i$, which can be used for the stabilizer code.


\section{Quantum-GRAND proposals}

We are now ready to construct GRAND-inspired decoding techniques to decode QRLCs. This common framework of quantum-GRAND (QGRAND), in the case of stabilizer codes, consists of the steps herein summarized:
\begin{enumerate}
	\item Generate a random QRLC encoding $U$, robust to the noise statistics $\mathcal N$. This can be efficiently done, for instance, by combining random 2-qubit Clifford unitaries (\cref{sect:encoding,sect:robust_encoding});
	\item Determine stabilizers of the chosen encoding by efficiently simulating the stabilizer circuit (\cref{sect:stabilizers});
	\item Perform syndrome measurement using all the minimal stabilizers (\cref{sect:stabilizer_code});
	\item Apply decoding using an iterative noise guessing recovery procedure, similar to the one in classical GRAND (\cref{sect:ClassicalGRAND}), by applying the inverse of the error pattern determined to the encoded quantum data.
\end{enumerate}

These steps can be implemented in slightly different manners, as will be described in the following subsections.

\subsection{The syndrome as membership test}

\begin{figure}[!ht]
	\centering
	\subfigure[]{\includegraphics[width=0.8\linewidth]{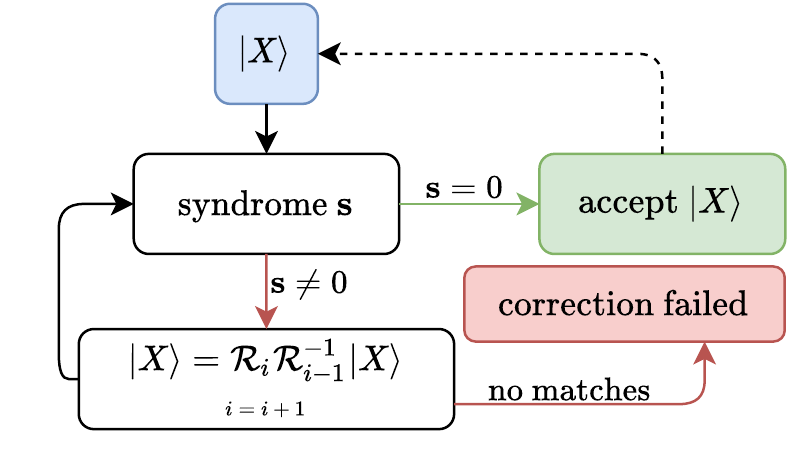}\label{fig:QGRAND_flowchart_a}}
	\subfigure[]{\includegraphics[width=0.8\linewidth]{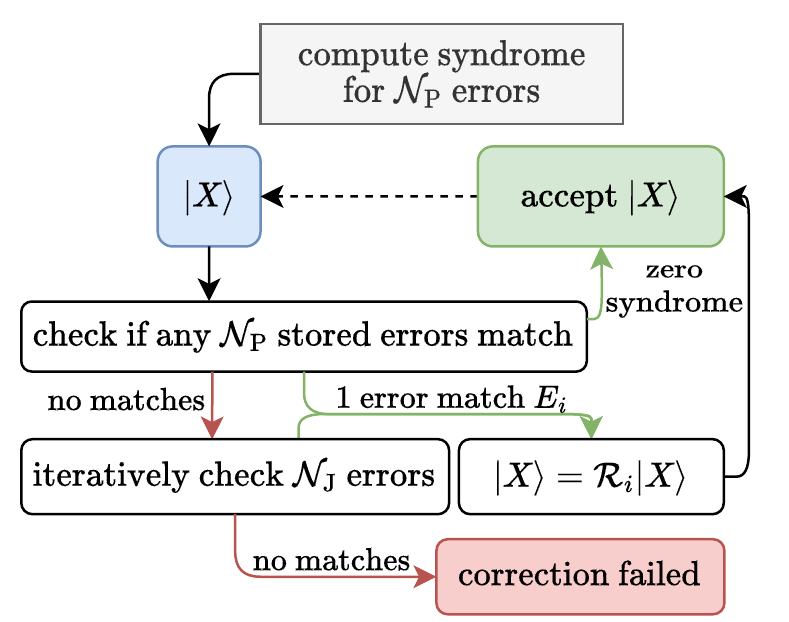}\label{fig:QGRAND_flowchart_b}}
	\subfigure[]{\includegraphics[width=0.8\linewidth]{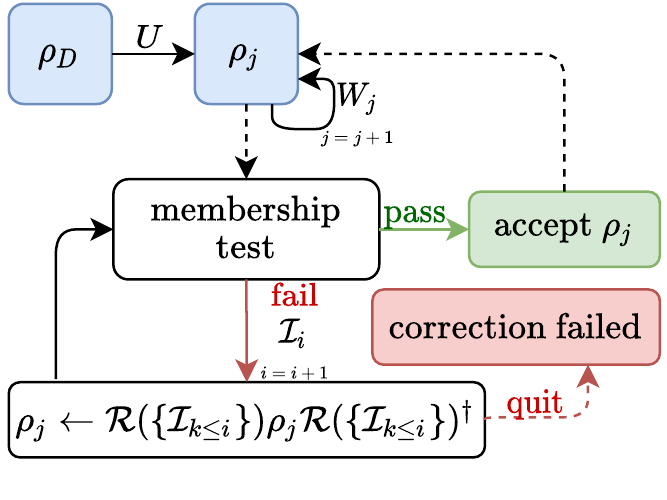}\label{fig:QGRAND_flowchart_c}}
	\caption{Flowchart of QGRAND, when (a) using the syndrome solely as a membership test, or (b) also as an aid for decoding. Going beyond stabilizer codes, (c) QGRAND may also be employed in a more general context.}
	\label{fig:QGRAND_flowchart}
\end{figure}

Syndrome measurement setups are a central building block in QEEC \cite{Swathi_Access_2022, LDPC_quasi-cyclic_stabilizer_TCOM_2018, Wang_TCOM_23}. As in classical GRAND, the syndrome measurement step of the stabilizer code may be considered simply as a membership test, accepting the quantum state if the syndrome $\mathbf{s}$ is zero, and rejecting the error-affected state if $\mathbf{s}\neq \mathbf{0}$ (see Fig. \ref{fig:QGRAND_flowchart_a}).
If the syndromes associated with each error pattern are not known \emph{a priori}, then it will be necessary to measure all $s=n-k$ minimal stabilizers to verify if no error has occurred. Since the encoding and subsequent stabilizers are random, it will take only $\sim 2$ stabilizer measurements, on average, to obtain a nonzero syndrome bit when an error actually occurred.

For a $(n,k)$ code, we do not necessarily require the measurement of the full syndrome to perform the GRAND iteration procedure. If we are not taking advantage of the information encoded in the syndrome, and wish to use it only as a membership test for the codebook, then we just need to check whether or not the syndrome is nonzero.

As the chosen encoding is random, so are its stabilizers. Consequently, each error pattern has a probability of $p=1/2$ of anticommuting (and thereby being detected) by each stabilizer, and this behavior is independent for each error pattern. Therefore, the average number of stabilizers measured until encountering a 1 bit, for some error pattern, is
\begin{align}
	\ev{C_{s}(p)} = &\, \sum_{i=1}^{s} i(1-p)^{i-1}p\\
	\Rightarrow\;\ev{C_{s}}= &\, \ev{C_{s}(1/2)} = 2-\frac{s+2}{2^{s}} < 2.\label{eq:S_half}
\end{align}
As we generally don't know \emph{a priori} whether or not an error has occurred at the first iteration, before the trial-and-error correction process commences, there is the possibility that we will need to measure all $s$ minimal stabilizers, in the case with probability $p_0$ where no error has occurred. In that case, the average number of stabilizer measurements required to determine whether an error has occurred or not will be
\begin{align}
	\ev{C_{s}(p_0)} =  p_0 s + (1-p_0)\ev{C_{s}}.
\end{align}
If an error is observed, then at each iteration $i>0$ of the trial-and-error correction attempt, the average number of measurements is $C_{s}(p_i / q_i)$, with $q_i \triangleq  1-\sum_{j=0}^{i-1}p_j$.

The average total number of iterations in the trial-and-error procedure, for $N$ error patterns, is $I \triangleq  \sum_{i=0}^N (i+1) p_i$.
Consequently, the average total number of stabilizer measurements until the procedure is completed is
\begin{align}
	\ev{C} = s + (I - 1)\ev{C_{s}} < s-2 + 2I.
\end{align}

A downside of this approach is that the error correction and syndrome extraction must be applied physically in the quantum circuit at each iteration. As this correction consists of applying $E_i^\dagger = E_i$ to the circuit, and the syndrome extraction step might have a non-negligible number of gates (see Fig. \ref{fig:stab_code}), then, for error-prone implementations, this approach may lead to a costly error correction process, as each application of correction gates adds error itself to the encoded state.

For the case without abandonment, the error correction process can fail solely when an observed error pattern has the same syndrome as an error pattern that was more likely to occur, given the noise statistics. This amounts to the scenario where the coset leader chosen for a syndrome is not the error that occurred. As previously alluded to in \cref{sect:ClassicalGRAND}, when abandonment is implemented, a second source of failure is the probability that an unlikely error occurs, beyond the errors that one wishes to correct. Nonetheless, as indicated in \cref{sect:robust_encoding}, $n$ should be chosen to ensure that this probability of failure is low.

\subsection{Syndrome decoding}

Alternatively, as long as partial or full access to the actual syndrome is available, one may use that information to improve the error correction process, while still taking advantage of the known noise statistics (see Fig. \ref{fig:QGRAND_flowchart_b}). There are several possible approaches along this route. For instance, one may rely on the classical computation of the syndromes associated with each relevant error pattern, both before and during the correction process.

To take advantage of syndrome decoding, let $\mathcal N_{\rm P} \subseteq \mathcal N$ be the set of error patterns for which we wish to precompute and store the error syndrome, and $\mathcal N_{\rm J}=\mathcal N \backslash \mathcal N_{\rm P}$ the error patterns for which we wish to compute the error syndrome on the fly, as necessary. We have $N = N_{\rm P} + N_{\rm J}$, with $N_{\rm P} \triangleq  |\mathcal N_{\rm P}|, N_{\rm J} \triangleq  |\mathcal N_{\rm J}|$.
Its $N_{\rm P} \times s$ syndrome matrix $M_{\rm S}$, which stores the $N_{\rm P}$ syndromes, each $s$-bits long, can be computed by the matrix product of the $N_{\rm P} \times 2n$ error matrix $M_{\rm E}$ and the $ S \times 2n$ parity check matrix $A$, through $M_{\rm S} = M_{\rm E} A^T$ \cite{djordjevic_quantum_2021}.
If $N_{\rm P}$ is small, this syndrome matrix can be classically stored in memory.

Additionally, the precomputation of the $\mathcal N_{\rm P}$ error patterns' syndromes enables checking \emph{a priori} whether or not our random linear code can distinguish between all the $\mathcal N_{\rm P}$ error patterns, lest it have more than one relevant error pattern with the same syndrome (cf. \cref{sect:uniform_at_random}). Therefore, we may try other random encodings until we are sure that one satisfies the requirements at hand, instead of relying on \eqref{eq:n_requirement}.

Given the random nature of the linear code, it takes, on average, $\log_2 N_{\rm P}$ bits to identify each of the $\mathcal N_{\rm P}$ error patterns, for a total of at most $\sim 2N_{\rm P}$ bits to store, using a binary tree.
To single out a particular error pattern in $\mathcal N$, we require, on average, $\log_2 (N+1)$ stabilizer measurements.
Note that $\log_2 (N+1)$ may be much less than $s$ bits in practice (see \eqref{eq:n_requirement}).
After $\log_2 (N_{\rm P}+1)$ stabilizer measurements, we are likely to rule out almost all error patterns in $\mathcal N_{\rm P}$. However, the error patterns in $\mathcal N_{\rm J}$ have not yet been ruled out. To be sure of which error actually occurred, $\log_2(N_{\rm P}+N_{\rm J}+1)$ measurements are needed.
For that, the syndromes of the other $\mathcal N_{\rm J}$ error patterns needed to be checked. These are error patterns that we still wish to correct, but whose syndrome we did not keep in memory due to memory constraints. These computations can be performed in parallel, and we only need to compare, and thus compute, on average, $<2$ bits among the measured syndrome bits, to check whether one of these error patterns is the measured one.

In the case with abandonment, it is possible that none of the $\mathcal N_{\rm P}$ or $\mathcal N_{\rm J}$ error patterns explain the observed syndrome, if the error that occurred was very unlikely and not worth correcting.

\subsection{Ordering the stabilizers}

With direct access to the syndrome information, it might be tempting to order the stabilizer measurements so that, at each iteration, there is maximum information gain. However, it can be shown that this costly computation results in negligible savings in the average number of measurements that need to be performed to determine the error (see \cref{sect:stabilizer_ordering}), when compared with a random ordering. Similarly, ordering the stabilizers to try and create a more efficient membership test would also result in minimal improvement in performance.

If $s \gg \log_2 (N+1) > H$, one may work with a smaller subset of the $s$ minimal stabilizers, since, on average, we only require $\log_2(N+1)$ measurements (in the highest entropy case) to determine the error. In this case, it is possible to choose stabilizers that have low weight or that are easier to implement, based on the hardware restrictions. However, this biased sampling may negatively affect performance. In any case, it should be noted that, if the encoding is chosen according to \eqref{eq:n_requirement}, then one should not have $s\gg H$.

\subsection{Adapting to changing noise statistics}

As the set of noise statistics $\mathcal N$ takes center stage in the decoding process, QGRAND may be adapted on the fly to suit a changing noise environment, or changing error tolerances. For the latter case, less (more) strict correction expectations can be achieved by discarding (including) some error patterns $E_i$ with low associated $p_i$. Similarly, for the former case, if $\mathcal N$ no longer accurately represents the new noise environment, a new $\mathcal N'$ may be considered, resulting in different errors having correction priority. 

If changing $\mathcal N$ is not sufficient, the code can still be quickly adapted by undoing the old encoding $U$ and implementing a new encoding $U'$, keeping in mind the expected performance indicated by \eqref{eq:n_requirement}, \eqref{eq:P_ineq}, and \eqref{eq:f_ineq}.  

\subsection{A general approach: beyond stabilized codes}

The QGRAND concept is not limited to stabilizer codes. A general high-level implementation of QGRAND-inspired decoders may be applied to other QECCs, consisting of the following steps, as depicted in Fig. \ref{fig:QGRAND_flowchart_c}:
\begin{enumerate}
	\item\textbf{Preprocessing/Encoding}: The initial quantum data, $\rho_D$, is preprocessed by $U$, possibly composed by ancilla qubits, unitary transformations, and projective measurements, into its encoded version $\rho_{j}$ (with $j=0$). $\rho_j$ enables the desired operations $W_j$ on the quantum data to take place, transforming the state into $\rho_{j>0}$. All of these states $\rho_j,\; j\geq 0$ should be robust to errors. At desired intermediate steps, the data is verified.
	\item\textbf{Data verification}: A \emph{membership test} is performed to verify if the encoded state $\rho_j$ has been corrupted by errors beyond the maximum tolerable error weight - \emph{fail} case - or if it is still in a state acceptable for further operations - \emph{pass} case. Depending on the implementation, the membership test may provide further information $\mathcal{I}_i$ about the failure, facilitating its subsequent correction. Note that the use of the membership test may itself modify $\rho_j$, as long as this modification simply aids the correction, and does not induce further significant errors in $\rho_j$.
	\item\textbf{Iterative correction attempt:} A recovery operation $\mathcal R$ is attempted on the state $\rho_j$. This operation is chosen based on the information $\mathcal{I}_k$ provided by the previous membership tests, and on the noise statistics $\mathcal N$ known about the setup \emph{a priori}. This verification should be reversible, so that subsequent iterations may undo the effects of failed correction attempts.
	\item\textbf{Error correction}: Steps 2 and 3 are repeated either until no feasible further recovery attempts are possible - leading to a failed correction - or until a membership test indicates that $\rho_j$ is now in an acceptable state, and the potential operations $W_j$ can resume.
\end{enumerate}

\section{Performance vs code rate trade-off} \label{sect:results}

\begin{figure}[t]
	\centering
	\includegraphics[width=0.88\linewidth]{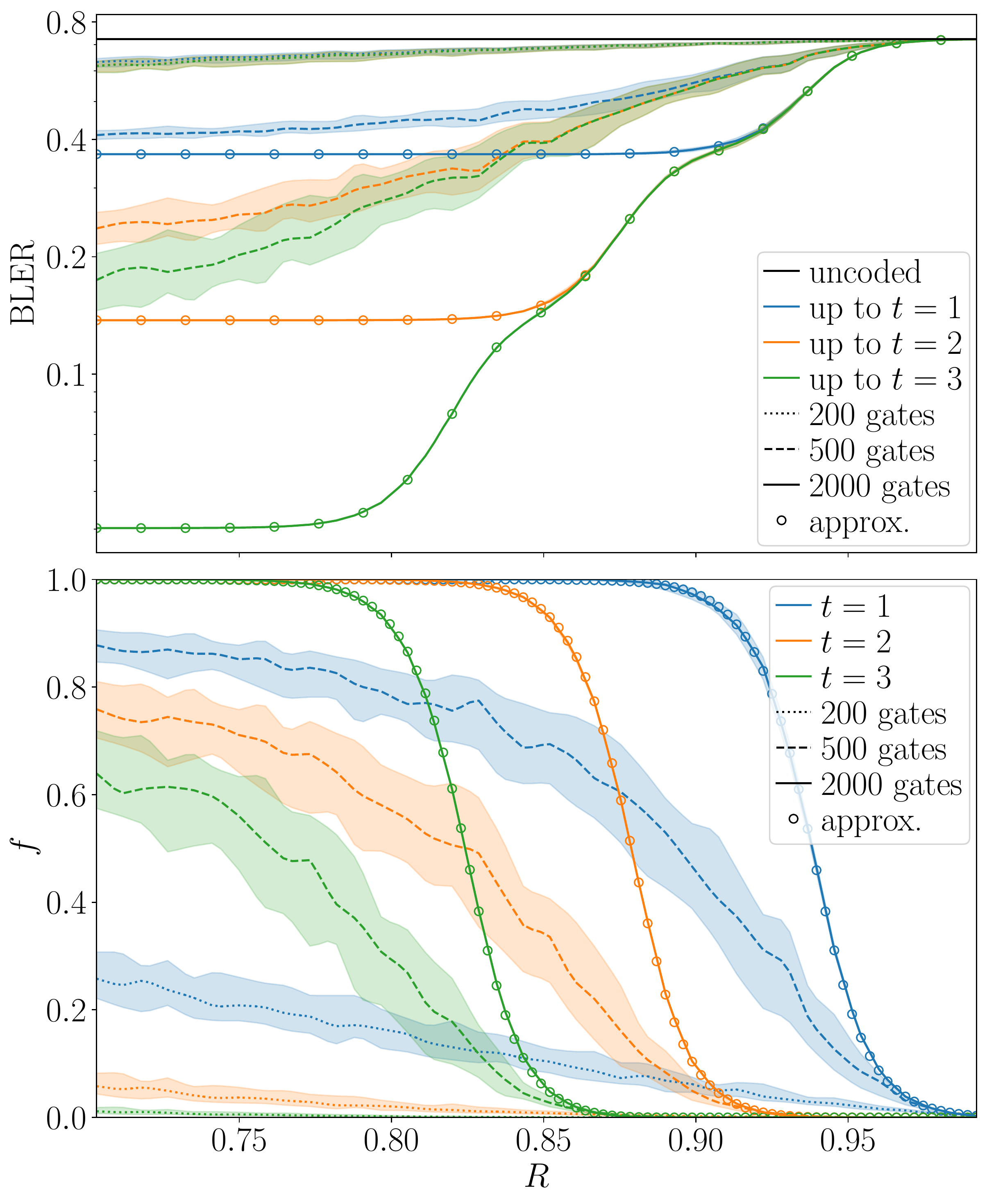}
	\caption{BLER and correctable error fraction $f$ for QRLCs with $n=128$, considering $p=10^{-2}$, when trying to correct error patterns of weight up to $t=3$. Top curves: BLER, with shaded areas marking an $80\%$ confidence interval. Bottom curves: fraction $f$ of error patterns of weight $t$ that can be corrected by the code, for a given code rate.
	}
	\label{fig:128}
\end{figure}
\begin{figure}[t]
	\centering
	\includegraphics[width=\linewidth]{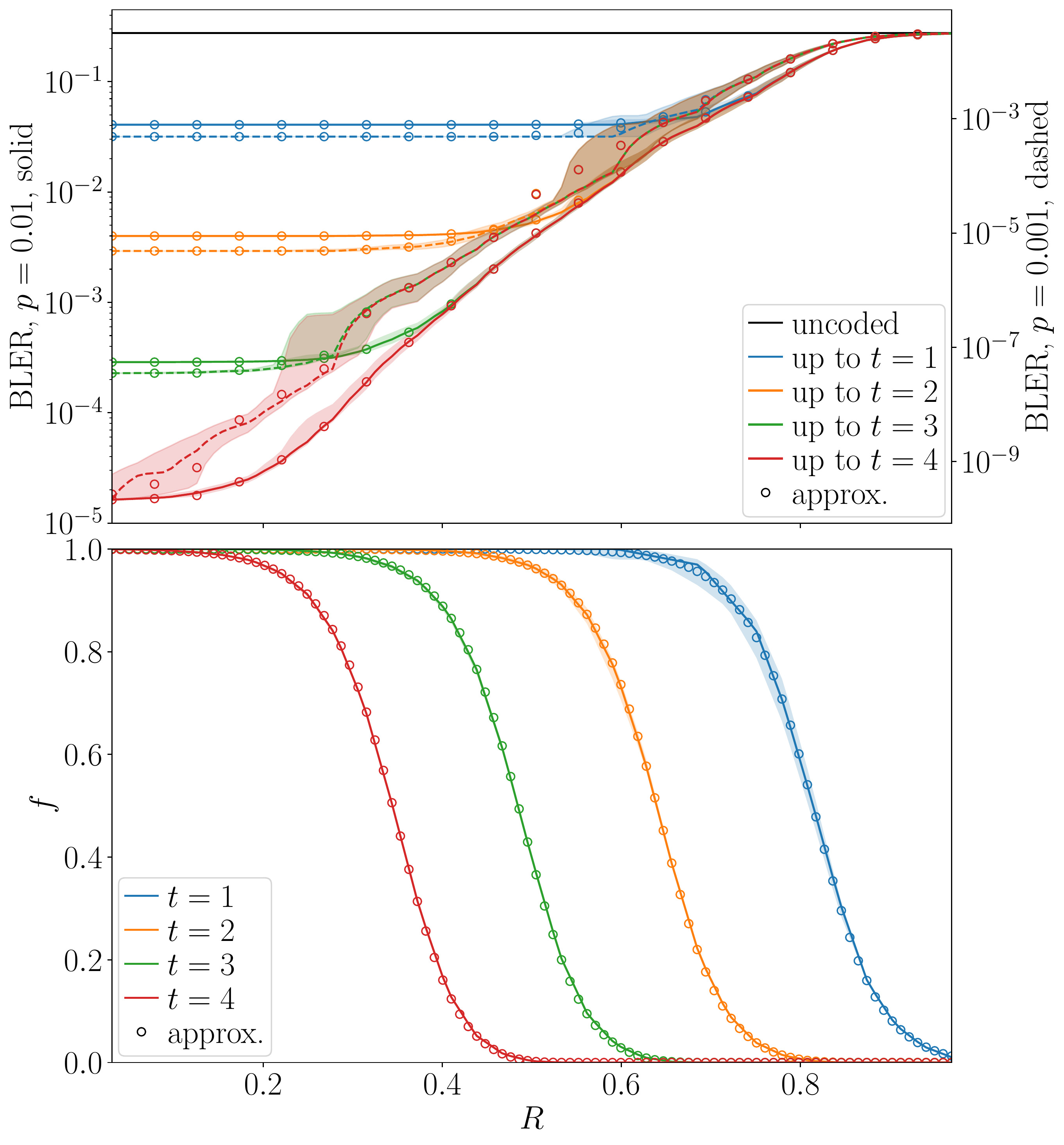}
	\caption{BLER and correctable error fraction for QRLCs with $n=32$, considering different $p$ values, when trying to correct error patterns of weight up to $t=3$. Top curves: BLER, with an overall behavior similar for both $p=10^{-2}$ and $p=10^{-3}$, but the lower $p$ enables 2 to 3 orders of magnitude lower error rate. Bottom curves: fraction $f$ of error patterns of weight $t$ that can be corrected by the code, for a given code rate. This value is independent of $p$.}
	\label{fig:32}
\end{figure}

As motivated in \cref{sect:robust_encoding}, the alteration of quantum states follows a noise model where a quantum state $\sigma$, in density matrix form, gets transformed into
\begin{equation}
	\rho = \sum_{i}^{N} p_i E_i \sigma E_i^\dagger,\quad 
	\text{with }E_0 = I.\label{eq:rho}
\end{equation}
After measuring some syndrome $\mathbf{s}$, and applying the associated correction (either directly, if possible, or by iteratively guessing), we obtain the corrected state. Unlike the classical case, where bit-flip errors definitely map a codeword to a different word, in the quantum case the error's actual effect will depend on the original quantum data $\ket \psi$, and it may go from doing nothing to mapping $\ket{\psi}$ to an orthogonal quantum state. As an example, consider $k=1$ and only the error $\bar Z$. If $\ket{\psi} = \alpha\ket 0 + \beta \ket 1$ (with $|\alpha|^2 + |\beta|^2 = 1$), and the affected state $\bar Z \ket \psi$ is not corrected, the decoded state's fidelity will be
\begin{equation}
	F = |\ev{\bar Z}{\psi}|^2 = ||\alpha|^2 - |\beta|^2|^2, 
\end{equation}
which may be any value from zero (if $|\alpha|=|\beta|$) to one (if $|\alpha|$ or $|\beta|$ equal one). As we wish to determine an estimate of the QGRAND's performance which is independent of the initial quantum data of interest, we may estimate a lower bound $F_{\rm min}$ for the fidelity of the corrected state, averaged over the noise probability distribution, for any given initial quantum state. A straightforward lower bound may be obtained by considering the worst-case scenario, where all errors in $\mathcal E$ (except $E_0=I$) map the initial quantum data to an orthogonal state, resulting in $F=0$ if an error occurs. Therefore, we have the lower bound
\begin{align}
	\ev{F} \geq F_{\rm min} &= \sum_{i\in \mathcal C} \epsilon_{i} \geq \sum_{i\in \mathcal C_{\rm non-degen.}} \epsilon_{i} = 1-\text{BLER},
\end{align}
with $\mathcal C$ the set of indices for error patterns that can be successfully corrected by the code. $\mathcal C_{\rm non-degen.}\subseteq \mathcal C$ considers only errors that are correctable thanks to a distinct syndrome, and not thanks to code degeneracy. $1-F_{\rm min}$ turns out to be further lower bounded by the block error rate (BLER) in classical error correction, with the difference resulting from code degeneracy. Since the probability of an error being degenerate roughly scales with $1/L=1/2^{2k}$ (cf. \cref{sect:uniform_at_random}), this possibility can be disregarded for moderate or high $k$. Nonetheless, by computing the BLER, our results slightly underestimate the codes' performance for very low values of $k$.

The error probability associated with quantum communications and quantum computation is typically in the order of $10^{-3} - 10^{-1}$ (\cite{PhysRevLett.124.110501,Yin2012,Huang2019}), higher than what is usually encountered in classical applications. For fault-tolerant computation, it is estimated that an error probability below $\sim 10^{-2}$ would be required \cite{faulttolerance}. To motivate the use of QGRAND for near-future applications, for this work, we consider error probabilities in this range.

Figures \ref{fig:128} and \ref{fig:32} show how both the block error rate (BLER, which is an upper bound to the infidelity in the quantum setting) and the fraction of error patterns of weight $t$ evolve as a function of the code rate for QRLCs with codeword lengths $n=128$ and $n=32$, respectively.

These results are obtained semi-analytically, as the QRLCs are constructed randomly, but their performance is computed exactly from each QRLCs' list of minimal stabilizers, represented by the parity check matrix $ A$. The implementation uses the Python package \emph{Qiskit}. For $n=128$, we compute 31 samples for each $k\geq 90$. For $n=32$, all $k$ are considered. For each sample, a QRLC with the indicated number of gates (it varies for $n=128$, and it is always 2000 for $n=32$) is constructed. Its performance is obtained by summing the probabilities $p_i$ of the most likely error patterns $E_i \in \mathcal E$ associated with each syndrome. These are the error patterns that can, in fact, be corrected by QGRAND (disregarding degeneracies). The syndrome of each error is efficiently computed as $\vb s_i = M_{E_i} A^T$, with $M_{E_i}$ the binary matrix representation of $E_i$. Only error patterns up to the desired weight $t$ are considered; thus $N=B_t$ as in \eqref{eq:B_t}.

As the code rate increases, the code manages to correct fewer and fewer error patterns, with an error rate approaching the case where no error can be corrected. Lower code rates eventually plateau for each weight $t$, as the codes can eventually correct all error patterns with that weight. As expected from Fig. \ref{fig:relative_distance}, if the encoding has too few $\mathcal C_2$ gates, it is not expressive enough to reach the performance of an ideal random code (marked by dots, \eqref{eq:f}), but, as more gates are used, the observed performance of these QRLCs matches it, qualitatively suggesting that GRAND's results for uniform-at-random codes may be applicable to quantum random encodings. Note that the variance of the code performance also decreases with an increasing number of gates, indicating that a good code is obtained with high probability.

\begin{figure}[t]
	\centering
	\includegraphics[width=0.95\linewidth]{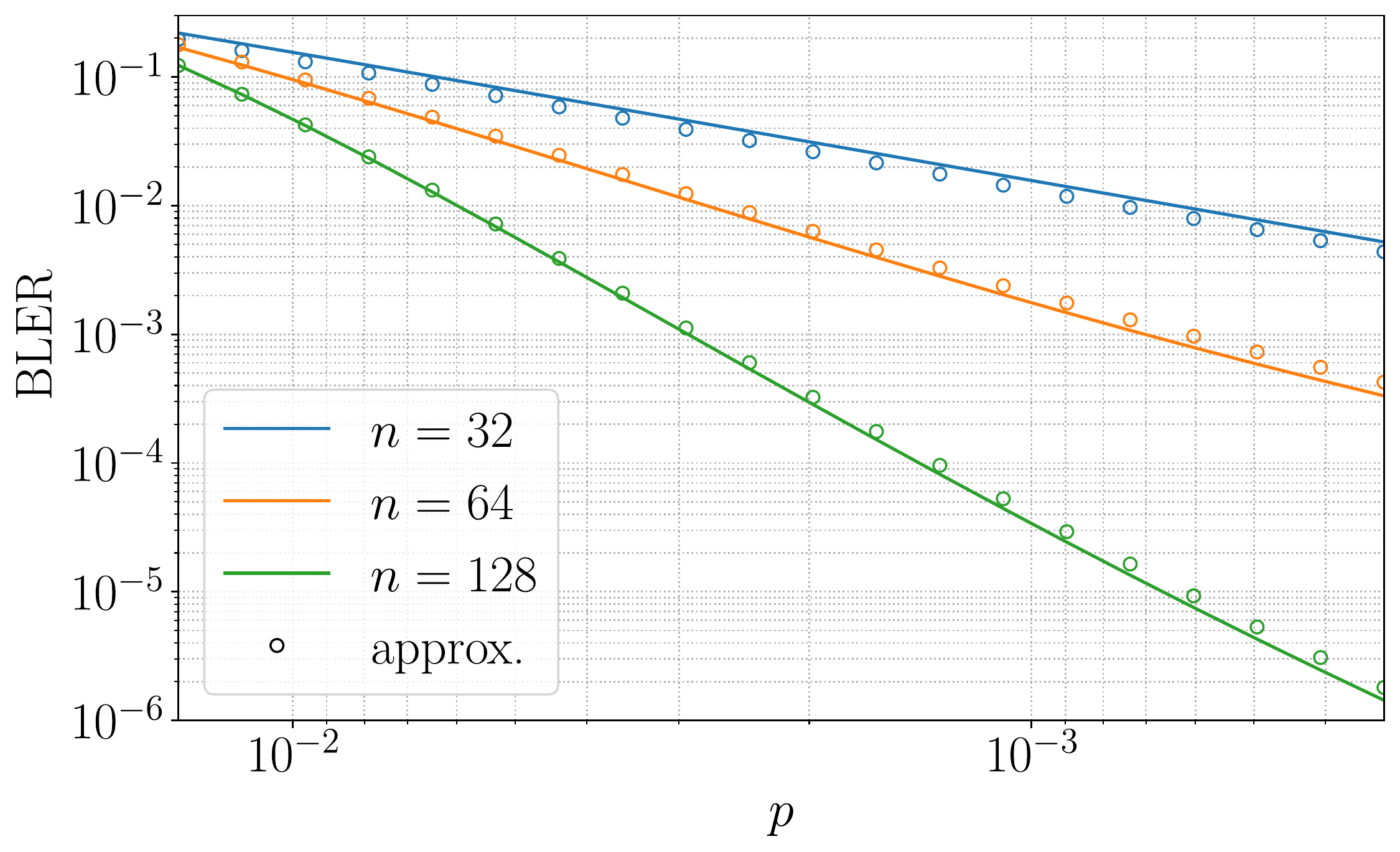}
	\caption{Codes' performance as a function of the probability $p$ of each qubit suffering an error for different code lengths $n$, with $R=0.7$. A Bernoulli noise model is used.}
	\label{fig:p_threshold}
\end{figure}

Approximating QRLCs by their ideal counterparts, we may estimate their performance's dependence on the noise probability. As seen in Fig. \ref{fig:p_threshold}, increasing $n$ improves the code's performance, with a lower $p$ observing greater improvements. The ideal approximation from \cref{sect:uniform_at_random} (dots) continues to provide a quick and good estimate of QGRAND's performance.  

\section{Conclusions and future directions}\label{sect:conclusions}

This work introduced a novel approach to quantum error correction by bringing together quantum random linear encoding and noise guessing decoding, considering a system model with channel errors only, as a first extension of the GRAND concept to the quantum realm.
QGRAND is particularly suitable for codes with short codewords, or with high code rate, as highlighted in \cite[sec. VI-B]{Access_Hanzo_Sep2023}. Given the difficulties of manipulating large quantities of qubits, this type of error protection seems particularly fit to the presently envisaged quantum technologies requiring error correction, such as the transmission of quantum entanglement, quantum repeaters, quantum memories, and quantum computing. 

It was first verified by numerical simulations that the construction leading to random quantum codes can be efficiently implemented, even outside the regime with theoretical guarantees.
Although we focus on the case of all-to-all qubit connectivity, this method can be easily generalized to more restrictive connectivities.
A possible approach would be to limit the 2-qubit gates composing the encoding to only connect pairs of qubits with a direct connection. 
Employing higher circuit depths to achieve similarly powerful encodings, and limiting the stabilizer weight \cite{Gullans_Krastanov_Huse_Jiang_Flammia_2021} would enable versatile practical implementations.

By taking advantage of the noise statistics, known from the medium at hand, an efficient and flexible decoding process is obtained. One may even consider a scenario where the syndrome measurements are not promptly available, and rather there is only access to a membership test, flagging whether the syndrome is zero or not. The QGRAND methodology is generalizable beyond stabilizer codes, and is suitable for time-variant noise statistics and changing setup requirements, updated on the fly.

In the quantum setting, we may consider different forms of abandonment. In the classical GRANDAB approach, the iterative noise guessing process is stopped after a chosen number of attempts, depending on the noise entropy. In the case where the syndrome information is being actively used, we may not only stop testing error patterns after a given threshold, but we may also stop measuring stabilizers after a given number of measurements, if the conducted measurements rule out all but very unlikely errors. In general, we expect the QGRAND approach to be directly amenable to the GRANDAB formalism. 

This work considered QECCs mostly for quantum communications and quantum memories, where the errors mostly arise during the quantum state transmission or storage. However, in contexts such as quantum computation, the errors during the encoding and decoding process themselves may be of the same order of magnitude as those encountered in the intended computations, and therefore, an important research question is to find how robust the whole encoding-decoding chain is when considering such errors in the system model. The first step should be assessing the impact of imperfect measurements when obtaining the syndrome, and how the binary decision they control (i.e., the membership test) is perturbed.
The application and efficiency of QGRAND (and GRANDAB) for practical applications with limited qubit connectivity, and also in the context of fault-tolerant systems, is the subject of ongoing work.

\appendices
\crefalias{section}{appendix}
\crefalias{subsection}{appendix}

\section{Stabilizer order} \label{sect:stabilizer_ordering}

For the syndrome decoding process, we may choose the order in which we apply the stabilizers. For a random ordering of the $s$ minimal stabilizers, it will take, on average, measuring $\sim \log_2 (N+1)$ stabilizers to identify which specific error pattern in $\mathcal E$ has occurred. If the syndrome is only being used as a membership test, then one needs to distinguish each error pattern in $\mathcal E \times \mathcal E$, which takes $\sim 2 \log_2 (N+1)$.

Instead of choosing the stabilizer ordering at random, we may try to choose the stabilizer with highest information gain, which, thanks to the random assignment of syndromes to error patterns, will likely correspond to the stabilizer with which the most error patterns anticommute.

\begin{figure}[!htpb]
	\centering
	\includegraphics[width=0.9  \linewidth]{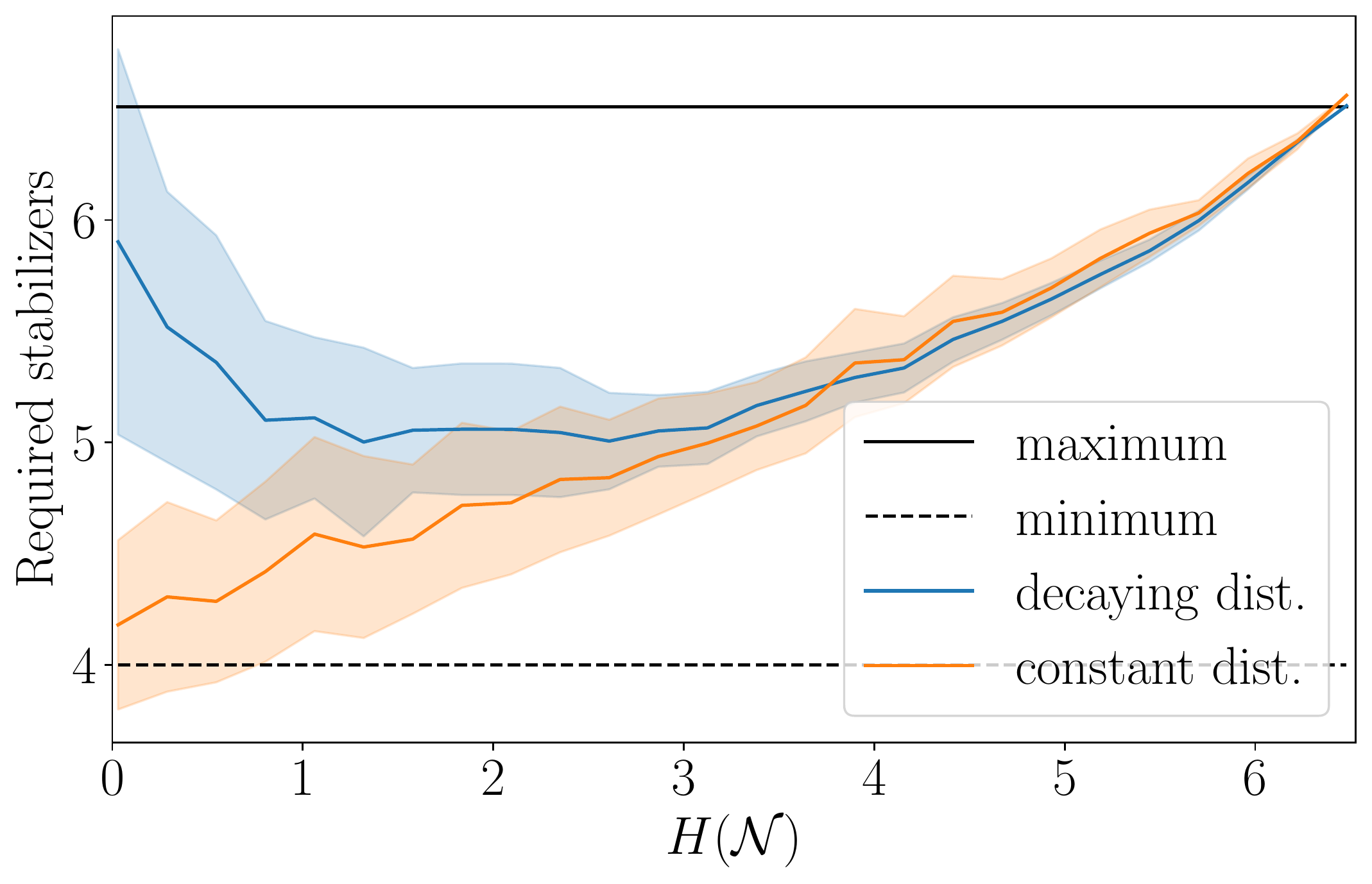}
	\caption{Average number of stabilizers required to determine the error that occurred, or if no error occurred at all, for different error distributions, as a function of the noise entropy $H(\mathcal N)$. The shaded region marks one standard deviation. Random $(30, 1)$ codes are used, and only error patterns of weight $t=1$ are considered.}
	\label{fig:decision_tree}
\end{figure}

Using this information, one may precompute a decision tree that maximizes the information gain at each measurement step. An example can be seen in Fig. \ref{fig:decision_tree}, where we see the average number of stabilizers required to determine the error that occurred, or if no error occurred at all, for different error distributions, as a function of the noise entropy $H(\mathcal N)$. The shaded region marks one standard deviation. Random $(30, 1)$ codes are used, and only error patterns of weight $t=1$ are considered. When all cases are equally likely and the entropy is maximal, the decision tree method will require about the same number of stabilizers, $\log_2(N+1)$ (black solid line), as the naïve method, where the stabilizers are chosen at random. For lower entropy distributions, only minor savings can be achieved. In the limit of zero entropy, where no errors can occur, fewer stabilizers are required (with the minimum given by the dashed line, computed using \eqref{eq:decision_tree_ordering}).
The decaying and constant distributions correspond to $\qty{p_i = \exp(-\alpha i)}$ and $\qty{p_i = \epsilon, p_0 = 1-N\epsilon}$, respectively, with both distributions normalized and $\alpha, \epsilon$ chosen so as to form a probability distribution with the indicated Shannon entropy $H (\mathcal N)$.

This approach might even be used to create a membership test requiring a smaller number of stabilizers.
However, here we show that such a procedure will lead to very minor savings in the necessary number of iterations, when compared with the straightforward random ordering.

Let $X_i$ be the distribution for the number of error patterns in $\mathcal E$ that anticommute with stabilizer $S_i \in \mathcal S_{\mathrm{min}}$. Then, approximately, $X_i\sim \text{Binomial}(N, 1/2)$, using the binomial distribution. For large $N$, $X_i$ can be approximated by the normal distribution $\text{Normal}(n/2, n/4)$.
The stabilizer that can detect the highest number of error patterns, will detect on average $Z \sim \max_i{X_i}$ error patterns. By Jensen's inequality, its expected value is upper-bounded by $\mu + \sigma \sqrt{2\ln s} = n/2 + \sqrt{(n/2)\ln s}$. Assuming that the stabilizer chosen for each iteration matches this upper bound and the $X_i$ distributions are independent, we have $N_j$ error patterns remaining at iteration $j$, on average, when starting with $N$ error patterns. The recursion relation is given by
\begin{equation}
	N_0 = N,\qquad N_{j+1} = \frac{N_j}{2} - \sqrt{\frac{N_j}{2}\ln(s-j)}.\label{eq:decision_tree_ordering}
\end{equation}

An approximate lower bound for the minimum number of measurements is given by $I$, such that $N_{I} \simeq 1$. This bound is not exact as the substitution of the binomial distribution for a normal distribution is only valid for large $N_j$, which necessarily cannot hold for $j\gg 1$. Nonetheless, it provides us with a simple rule of thumb to check whether the ordering approach can potentially have large savings in measurements. Using $A_j \triangleq  \sqrt{N_j / 2}$, we have
\begin{equation}
	A_{j+1} = \frac{A_j}{\sqrt{2}}\sqrt{1-\frac{1}{A_j}\sqrt{\ln(s-j)}}
	\simeq \frac{A_j}{\sqrt{2}}-\frac{\sqrt{\ln s}}{2\sqrt{2}}.
\end{equation}

\begin{figure}[t]
	\centering
	\includegraphics[width=0.9 \linewidth]{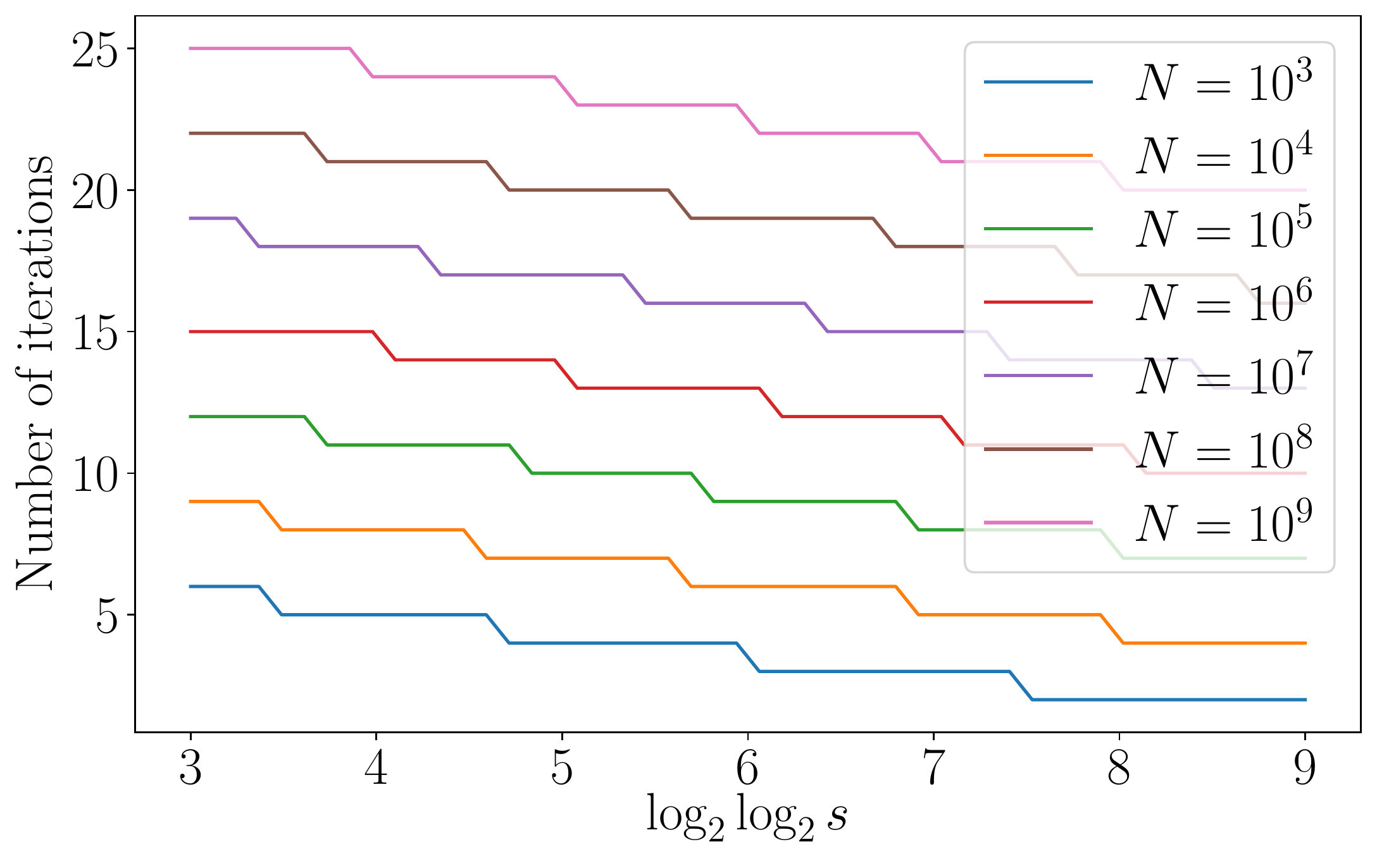}
	\caption{Expected number of iterations, using the decision tree method, for large $N$ and $s$. Using $I = a \log_2 N - b \log_2 \log_2 s - c$, a least squares fit obtains the parameters $(a,b,c)=(0.96, 0.91, 0.84)$, which is approximately in agreement with \eqref{eq:decision_tree_savings}.}
	\label{fig:decision_tree_savings}
\end{figure}

By solving this recursion relation, it can be shown that the number of iterations required will be, approximately, for large $N$ and $s$,
\begin{align}
	I \simeq \log_2 N - \log_2\log_2 s - 1\label{eq:decision_tree_savings}.
\end{align}
Fig. \ref{fig:decision_tree_savings} provides numerical confirmation of that.
Suppose that our setup follows \eqref{eq:n_requirement}, so that $s \propto \log_2 N$. As $N$ increases, we expect no significant increase in savings for the number of iterations, when compared with the approach of picking stabilizers at random, which takes $\sim\log_2 N$ iterations.
Thus, it is not justifiable, in practice, to compute a syndrome table to construct a decision tree leading to the highest information gain per iteration, as the gains are negligible.

\section*{Acknowledgments}
Francisco Monteiro is grateful to Prof. Frank Kschischang (University of Toronto) for discussions on noise-guessing decoding, and to Dr. Ioannis Chatzigeorgiou (Lancaster University) for discussions on random linear codes and noise-guessing decoding. All authors are thankful to Dr. Bill Munro (NTT Basic Research Labs, Japan) and Prof. Kae Nemoto (National Institute of Informatics, Japan) for insightful discussions about QGRAND.

\bibliographystyle{IEEEtran}
\bibliography{qgrand_v3.bib}

\vskip 0pt plus -1fil
\begin{IEEEbiography}[{\includegraphics[width=1in,height=1.25in,clip,keepaspectratio]{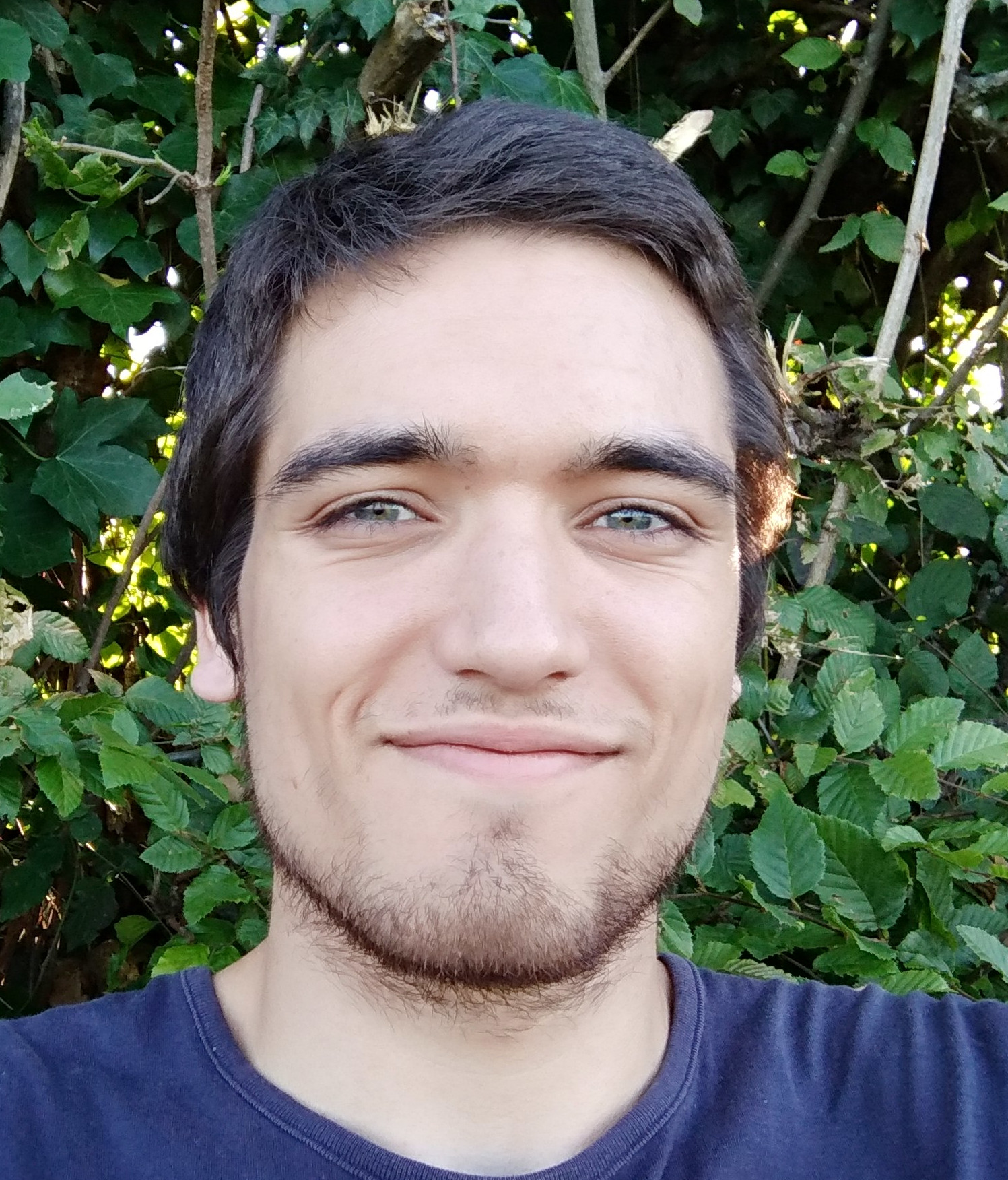}}]{Diogo Cruz} obtained his BSc and MSc degrees in Physics Engineering from Instituto Superior Técnico (IST), University of Lisbon, Portugal. He is currently a PhD student in Physics, also at IST, and is a researcher at Instituto de Telecomunicações, Lisbon, Portugal. He was a Calouste Gulbenkian Scholar in 2018/2019.
\end{IEEEbiography}

\vskip 0pt plus -1fil
\begin{IEEEbiography}[{\includegraphics[width=1in,height=1.25in,clip,keepaspectratio]{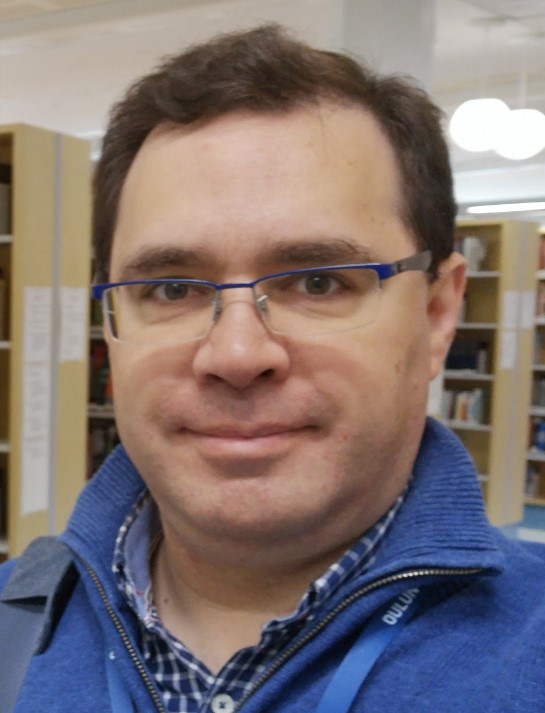}}]{Francisco A. Monteiro} (M'07) is Assistant Professor in the Dep. of Information Science and Technology at Iscte - University Institute of Lisbon, and a researcher at Instituto de Telecomunicações, Lisbon, Portugal. He holds a PhD from the University of Cambridge, UK, and the Licenciatura and MSc degrees in Electrical and Computer Engineering from IST, University of Lisbon, where he also became a Teaching Assistant. He held visiting research positions at the Universities of Toronto (Canada), Lancaster (UK), Oulu (Finland), and Pompeu Fabra (Barcelona, Spain). He has won two best paper prizes awards at IEEE conferences (2004 and 2007), a Young Engineer Prize (3rd place) from the Portuguese Engineers Institution (Ordem dos Engenheiros) in 2002, and for two years in a row was a recipient of Exemplary Reviewer Awards from the IEEE Wireless Communications Letters (in 2014 and in 2015). He co-edited the book ``MIMO Processing for 4G and Beyond: Fundamentals and Evolution'', published by CRC Press in 2014. In 2016 he was the Lead Guest Editor of a special issue on Network Coding of the EURASIP Journal on Advances in Signal Processing. He was a general chair of ISWCS 2018 - The 15th International Symposium on Wireless Communication Systems, an IEEE major conference in wireless communications.
\end{IEEEbiography}

\vskip 0pt plus -1fil
\begin{IEEEbiography}[{\includegraphics[width=1in,height=1.25in,clip,keepaspectratio]{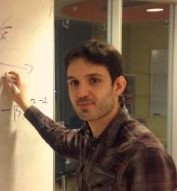}}]{Bruno C. Coutinho} obtained his PhD in Physics from Northeastern University, USA, in 2016, and both his BSc. and MSc. in Physics, from the University of Aveiro, Portugal, in 2009 and 2011, respectively. Since 2017 he is with the Physics of Information and Quantum Technologies Group, at Instituto de Telecomunicações, initially as a postdoc and later as a Research Fellow.
\end{IEEEbiography}

\end{document}